\documentclass[lineno]{jfm}

\usepackage{graphicx}
\usepackage{newtxtext}
\usepackage{newtxmath}
\usepackage{natbib}
\usepackage{float}
\usepackage{comment}
\usepackage{subcaption}
\usepackage{booktabs}
\usepackage{amsmath}
\usepackage{amssymb}
\usepackage{bbm}
\usepackage{bbold}

\usepackage{hyperref}
\hypersetup{
    colorlinks = true,
    urlcolor   = blue,
    citecolor  = black,
}

\newcommand{\RomanNumeralCaps}[1]
\linenumbers

\shorttitle{Data-driven equivariant LES model of \textcolor{black}{two-dimensional} turbulence}

\title{Physics-informed data-driven inference of an interpretable equivariant LES model of {\color{black}two-dimensional} incompressible fluid turbulence}

\author{
  Matteo Ugliotti\aff{1}\thanks{These authors contributed equally and share first authorship.}, Brandon Choi\aff{1}\footnotemark[1], Mateo Reynoso\aff{1}, Daniel R. Gurevich\aff{2},
 \and Roman O. Grigoriev\aff{1}\corresp{\email{roman.grigoriev@physics.gatech.edu}}}

\affiliation{\aff{1}School of Physics, Georgia Institute of Technology, Atlanta, GA 30332, USA
\aff{2}Department of Mathematics, University of California, Los Angeles, CA 90095, USA}

\begin{document}
\maketitle

\begin{abstract}
{\color{black}Restrictive phenomenological assumptions limit the accuracy and generality of subgrid-scale models of fluid turbulence, especially in flows where coherent structures and bidirectional energy transfer play an important dynamical role.
This paper introduces a symbolic, physically-informed data-driven subgrid-scale closure for two-dimensional incompressible turbulence.} {\color{black}The closure is inferred from filtered direct numerical simulation data using symmetry-constrained sparse regression, together with dimensional-consistency constraints on the inferred coefficients.} While the model is inferred using LES-style spatial coarse-graining, its structure is {\color{black}found to be} more similar to RANS models, as it employs an additional field to describe {\color{black}the dynamics of unresolved} scales. {\color{black}For the range of two-dimensional benchmark flows considered here, the inferred model substantially improves upon widely used LES models, yielding accurate predictions of subgrid-scale stresses, local and global fluxes between different scales, and other key quantities in both {\em a priori} and {\em a posteriori} benchmarks.}
\end{abstract}



\section{Introduction} \label{sec:introduction}
{\color{black}The objective of large eddy simulation (LES) is to provide an accurate and general description of the evolution of turbulent fluid flows at high Reynolds numbers when the cost of direct numerical simulation (DNS) becomes computationally prohibitive. Applying a spatial filter to the Navier--Stokes equations yields an evolution equation for the filtered velocity field $\bar{\bf u}$ which features an additional closure term $-\nabla\cdot\tau$, where the subgrid-scale (SGS) stress tensor $\tau_{ij}$ represents the effect of unresolved scales on the resolved dynamics.
The central modeling problem in LES is therefore to close the set of equations by expressing the SGS stress tensor (or its divergence) in terms of quantities that can be resolved on the computational grid.}

{\color{black}The majority of LES closures typically assume that the SGS stress tensor can be represented as a deterministic function of the filtered velocity $\bar{\bf u}$ and/or its derivatives. A particularly restrictive example of this is the Boussinesq assumption, which relates the trace-free symmetric part of the SGS stress to the resolved strain rate,
$\tau_{(ij)}=2\nu_e\nabla_{(i}\bar{u}_{j)}$, where $\nu_e$ is a scalar eddy viscosity and parentheses denote the trace-free symmetric component of a rank-two tensor. Popular models of three-dimensional turbulence such as the Smagorinsky model \citep{smagorinsky1963} and its dynamic version \citep{lilly_proposed_1992,germano_dynamic_1991} all use this assumption. Such models are classified as \textit{functional models} and are engineered to optimize the net energy dissipation. In contrast, \textit{structural models} seek to approximate the spatiotemporal structure of the SGS stress tensor itself, typically via formal series expansions or similarity arguments.} Representative examples include the Clark or nonlinear/gradient model (NGM) \citep{leonard1975,clark1979} and the similarity model \citep{Bardina1980}. Hybrid approaches combine elements of both classes and include models such as the dynamic nonlinear mixed model \citep{vreman_large-eddy_1996} and the dynamic mixed model \citep{Bardina1980}.

{\color{black}The present work focuses on building an LES model of two-dimensional incompressible turbulence in the bulk. Our broad objective here is to illustrate how a sufficiently general SGS model of turbulence can be inferred from DNS data representing a broad variety of flows without leveraging specific assumptions that may or may not hold for a specific flow. 
Two-dimensional turbulence is chosen due to its phenomenology that is distinct from its three-dimensional counterpart. In particular, the presence of pronounced coherent structures, inverse energy cascades, and significant local backscatter that are characteristic of two-dimensional turbulence all pose significant challenges for LES models. Therefore, two-dimensional turbulence presents itself as a useful testbed for SGS modeling in general.
 }

{\color{black}For two-dimensional turbulence, it is common to model the closure term using stochastic parameterizations \citep{berner_stochastic_2017} which describe the effect of unresolved scales rather than the resolved scales themselves, as is commonly done in, say, Reynolds-averaged Navier Stokes (RANS) models. Stochastic closures are particularly popular in weather and climate modeling, where forcing is often at small scales that are not explicitly resolved but do feed energy back into resolved scales. In this context, deterministic closures are generally found to under-represent stochastic variability. This finding is hardly surprising given that existing deterministic closures generally fail to accurately describe backscatter \citep{moser_statistical_2021}. }


Recent progress in machine learning (ML) has enabled \textcolor{black}{the inference of explicit parameterizations of closures from DNS data.} The most common approach is based on the application of the Cayley--Hamilton theorem to express the tensor $\tau_{ij}$ as a finite sum of linearly independent tensor basis functions, with coefficients given by scalar functions of a finite number of invariants of the \textcolor{black}{resolved velocity gradient }$\nabla_i\bar{u}_j$ \citep{pope_more_1975}. While this representation is formally complete, inferring the functional form of each coefficient for all of the invariants is extremely challenging. Consequently, practical implementations typically rely on truncating the expansion or restricting the number of invariants retained. Examples include sparse regression approaches such as random forest regression \citep{ling2016} and sequential thresholding ridge regression \citep{schmelzer2018}, and symbolic regression methods based on genetic expression programming (GEP) \citep{weatheritt2017,reissmann2021}.

Sparse regression has also been used to learn the explicit functional form of the closure term, component by component, in terms of the derivatives of the resolved fields, but without relying on the Cayley-Hamilton theorem. For instance, relevance vector machines (RVM) have been used to infer a quadratic parameterization resembling NGM for oceanic flows \citep{zanna2020} and two-dimensional turbulence \citep{jakhar2024}. A hybrid sparse/symbolic regression algorithm was employed to identify a (scalar) closure in a two-layer model of quasi-geostrophic turbulence \citep{ross2023} which involves higher-order derivatives of the large-scale variables (here, velocity and potential vorticity). However, none of these approaches ensure that the inferred closure transforms correctly under rotations or yields stable evolution \citep{jakhar2024}.

\textcolor{black}{A second class of ML approaches uses deep learning (DL) to yield an implicit, neural network-based parameterization. Such models have been applied to two-dimensional turbulence \citep{maulik2019,kochkov2021}, three-dimensional turbulence \citep{beck2019}, quasi-geostrophic flows \citep{bolton2019}, advection--diffusion problems \citep{frezat2021}, and MHD turbulence \citep{karpov2022}. Relevant physics, including symmetry, can be incorporated into DL parameterizations through hard or soft constraints, leading to improved performance \citep{frezat2021,guan2023,pawar2023}.} While DL-based approaches rely on less stringent assumptions, e.g., that the closure can be expressed solely in terms of the resolved variables, resulting parameterizations lack interpretability and have poor generalizability. Moreover, while DL parameterizations tend to reproduce the statistics well, they generally do not improve short-term predictions \citep{ross2023} and are commonly unstable \citep{guan2022,ross2023,pedersen2023}.

{\color{black}The specific aim of the present paper is to construct an explicit symbolic (and hence fully interpretable) SGS model of two-dimensional incompressible turbulence using a physics-informed data-driven procedure. Specifically, we use the SPIDER framework \citep{gurevich2024} which constrains the functional form of the individual governing equations making up the model using irreducible representations of the symmetry group generated by rotations and reflections. This model inference approach is fully data-driven for a given hypothesis space: the observables and the symmetry group are specified manually, while the functional form of the governing equations is inferred automatically from filtered DNS data.}


The paper is organized as follows. Section \ref{sec:methods} introduces our methodology. The results are presented in Section \ref{sec:results}. {\color{black}The structure of the inferred model and the limitations of the model and the modeling framework are} discussed in Section \ref{sec:discussion}. Finally, our conclusions are presented in Section \ref{sec:conclusions}. 


\section{Methods}\label{sec:methods}

\subsection{Governing Equations and Coarse-Graining}\label{section: Governing Equations}

We consider incompressible flows of Newtonian fluids described by the Navier-Stokes equation and the divergence-free condition 
\begin{subeqnarray} \label{eq:gov_eq}
    \partial_t u_i + u_j \nabla_j u_i &=& -\rho^{-1}\nabla_i p + \nu \nabla^2 u_i -\gamma u_i + f_i,\label{eq:ns}\\
    \nabla_i u_i &=& 0,\label{eq:div}
\end{subeqnarray}
where $\rho$ is the density, $\nu$ is the kinematic viscosity, $\gamma$ is the Rayleigh friction coefficient, $f_i$ is the forcing, and Einstein implicit summation notation is used. 
Since we are primarily interested in modeling turbulence far from any boundaries, we consider flows on a two-dimensional domain $\Omega$ of size $\ell_d\times\ell_d$ with periodic boundary conditions (and $\ell_d=2\pi$).

\begin{figure}
    \centering
    \begin{subfigure}{0.33\textwidth}
        \centering
        \includegraphics[width=\textwidth]{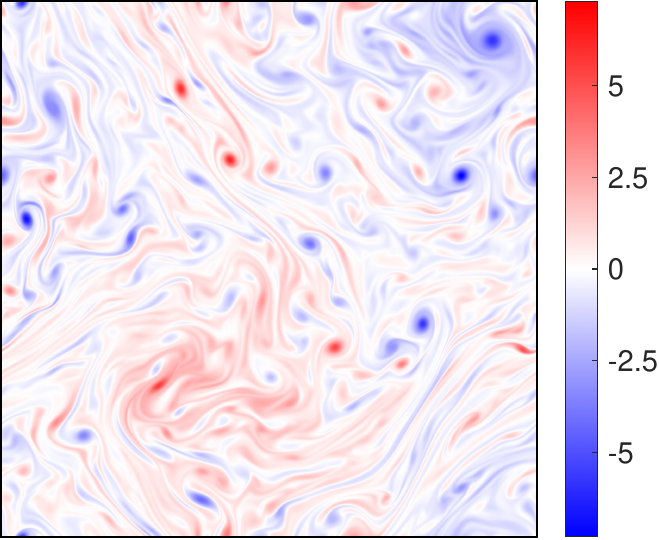}
        \caption{}
    \end{subfigure}
    \begin{subfigure}{0.32\textwidth}
        \centering
        \includegraphics[width=\textwidth]{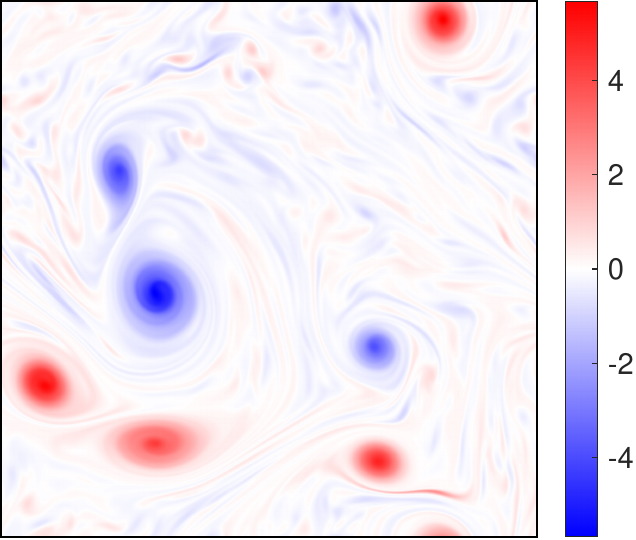}
        \caption{}
    \end{subfigure}
    \begin{subfigure}{0.33\textwidth}
        \centering
        \includegraphics[width=\textwidth]{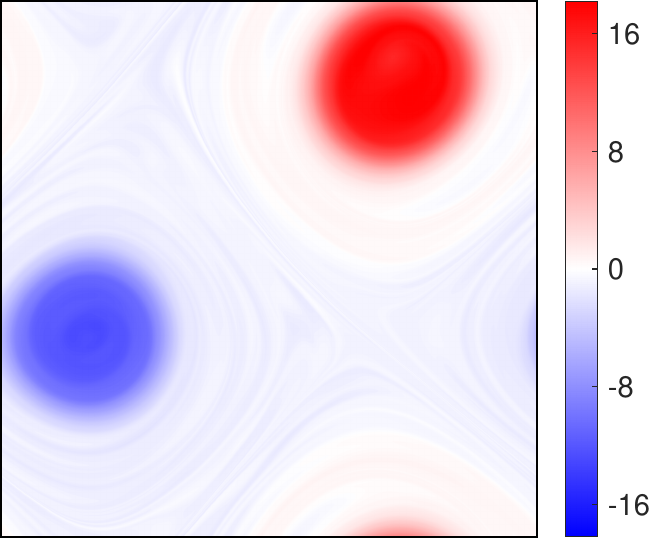}
        \caption{}
    \end{subfigure}

    \caption{
    Initial conditions for the three flows F1 (a), F2 (b), and F3 (c). \textcolor{black}{Vorticity field $\omega$ is shown.} 
    }
    \label{fig:IC_snapshots}
\end{figure}

Our training data was generated from DNS using initial conditions representative of three characteristic regimes of two-dimensional turbulence: forced, statistically stationary turbulence with direct or inverse cascade and transient freely decaying turbulence. The corresponding vorticity fields are shown in \autoref{fig:IC_snapshots} and feature coherent structures (vortices and vorticity filaments) with a characteristic length scale \textcolor{black}{$\ell_i$ (to be defined in Section \ref{section:model}) varying from $\ell_d/2$ to $\ell_d/30$.} For each of the three initial conditions, the flow was evolved for $O(1)$ eddy turnover time (here and below defined as $T_e = 1/ \omega_{\rm rms}$ computed using the initial condition) without forcing or Rayleigh friction and with the kinematic viscosity {\color{black}ranging from $10^{-4}$ to $10^{-6}$. The corresponding Reynolds numbers $\mathrm{Re} = \ell_d u_{\rm rms}/\nu$ are in the range from $2\times 10^3$ to $3\times 10^6$. Fully-resolved numerical solutions describing these flows were obtained 
using a pseudospectral solver with $2/3$ dealiasing. The detailed summary of parameters is provided in Table \ref{tab:apriori_flow_configurations}.} Additional details on our DNS are given in Appendix \ref{Appendix: Numerics}.

\begin{table}\color{black}
\centering
\setlength{\tabcolsep}{5pt}
\renewcommand{\arraystretch}{1.15}
\begin{tabular}{@{}lccc@{}}
\toprule
& F1 & F2 & F3 \\
\midrule
Regime
& Inverse cascade
& Free decay
& Direct cascade \\

$\nu$
& $10^{-4}$, $10^{-5}$, $10^{-6}$
& $10^{-4}$, $10^{-5}$, $10^{-6}$
& $10^{-4}$, $10^{-5}$, $10^{-6}$ \\

$\mathrm{Re}$
& $3\times10^3$ -- $3\times10^5$
& $2\times10^3$ -- $2\times10^5$
& $3\times10^4$ -- $3\times10^6$ \\

$\ell_i$
& $\ell_d/30$
& $\ell_d/4$
& $\ell_d/2$ \\


Forcing
& None
& None
& None \\

DNS grid at $\nu=10^{-6}$
& $2048^2$, $4096^2$, $4096^2$
& $2048^2$, $4096^2$, $4096^2$
& $2048^2$, $4096^2$, $4096^2$ \\


Figures
& \shortstack{\ref{fig:IC_snapshots}(a), \ref{fig:apri_corr}(a),
\ref{fig:apriori_fluxes}, \ref{fig:r_corr}, \ref{fig:lcr_split}}
& \shortstack{\ref{fig:IC_snapshots}(b),
\ref{fig:apri_corr}(b), \ref{fig:r_corr}}
& \shortstack{\ref{fig:IC_snapshots}(c),
\ref{fig:apri_corr}(c), \ref{fig:r_corr}} \\
\bottomrule
\end{tabular}
\caption{\color{black}Summary of data sets used for model inference and {\em a priori} benchmarks.}
\label{tab:apriori_flow_configurations}
\end{table}

Our subgrid-scale model was constructed using filtered DNS (FDNS) data, with the filter scales $\Delta$ chosen to lie between the fine (DNS) grid resolution and the length scale defined by $\ell_i$. We define the filter operator, acting on any tensor field $\phi$ using
\begin{equation} \label{filteroperator}
    \bar{\phi}(\textbf{x}) \equiv \int G(|{\bf x}-{\bf x}'|/\Delta)\phi({\bf x}') d\textbf{x}',
\end{equation}
where the kernel $G$ has width $\Delta$. 
Filtered variables are assumed to be resolved on a coarse (LES) grid,
unlike the fluctuations $\phi' \equiv \phi-\bar{\phi}$. 
The kernel $G$ should satisfy a number of properties, primarily locality in both real and spectral space, as to minimize the Gibbs phenomenon; the popular Gaussian kernel serves this purpose well \citep{pope_turbulent_2000,sagaut_large_2006,vreman_realizability_1994}. We further project the filtered field onto the coarse grid -- an operation mathematically equivalent to filtering again with a sharp spectral filter. This additional projection is necessary for data-driven discovery of accurate SGS models in the form of partial differential equations (PDEs), as the grid resolution strongly impacts {\color{black}higher-order} spatial derivatives.

Applying the combined filtering operator to the system of equations \eqref{eq:gov_eq} yields the governing equation for the filtered (resolved) velocity and pressure fields
\begin{subequations} \label{eq:f_gov_eq}
\begin{align}
    \partial_t \bar{u}_i + \bar{u}_j \nabla_j \bar{u}_i &= -\rho^{-1}\nabla_i \bar{p} + \nu \nabla^2 \bar{u}_i - \gamma \bar{u}_i + \bar{f}_i - \nabla_j\tau_{ij},\label{eq:f_ns}\\
    \quad \nabla_i \bar{u}_i&=0. \label{eq:f_divu}
\end{align}
\end{subequations}
The closure term $-\nabla_j\tau_{ij}$ is written in terms of the subgrid stress tensor $\tau_{ij} \equiv \overline{u_iu_j} - \bar{u}_i\bar{u}_j$. Note that the pure Gaussian filter is invertible, so ${\bf u}$ (and therefore $\tau$) can in principle be expressed in terms of $\bar{\bf u}$ alone in the continuum limit. However, on a discrete grid, the invertibility is lost and so there is no reason to expect that $\tau$ can be expressed solely in terms of the resolved {\color{black}velocity}, as is commonly assumed.

\subsection{Physically-informed data-driven equation inference}\label{section: SPIDER}

{\color{black}We aim to construct an SGS model in the form of a system of (nonlinear) partial differential equations. We also impose the requirement that the SGS model should inherit the symmetries of the physical space, i.e., the model must be symmetry-equivariant; in this case, it means that the model form should be independent of the position and orientation of the coordinate system, up to a group transformation.
The procedure for discovering an equivariant and dimensionally consistent model consists of four steps: (1) choosing the resolved observables, (2) constructing a sufficiently expressive equivariant symbolic search space, (3) using data to identify parsimonious functional relations between the resolved variables and/or their derivatives, and (4) uncovering a dimensionally consistent relation between the coefficients that appear in these relations and the parameters of the underlying problem, e.g., $\nu$ and $\Delta$.



The first step is to specify the variables, i.e., the fields describing the large and small scales. Large scales are naturally described by the coarse-grained original fields, i.e., $\bar{u}_i$} and $\bar{p}$. There is no general way to identify the SGS variables describing small scales {\it a priori}. However, formal analysis can help with this choice, as will be illustrated below. Given these observables, we accomplish steps (2) and (3) by using SPIDER, a data-driven equation inference framework for spatiotemporal data that automatically constructs symmetry-equivariant search spaces and identifies the parsimonious relations they contain \citep{gurevich_learning_2024}. Specifically, we use the Python implementation PySPIDER \citep{PySPIDER}.

SPIDER requires the relevant symmetry group to be specified; in our case, it is the Euclidean group $E_2$, which includes translations, rotations, and reflections. SPIDER also requires specifying how the chosen variables transform; we use tensor representations. Within each tensor representation, the search spaces are constructed automatically by taking all possible contractions between tensor products of the variables and their derivatives. This formal procedure generates infinitely many terms, so in practice the search space is truncated by term complexity, here the number of tensor indices---before contraction---appearing in a candidate term. The resulting terms are then sorted into libraries $\mathcal{L}_m$ corresponding to irreducible representations $m=1,2,\cdots$ of the specified symmetry group. For a fixed set of irreducible representations, here limited by the highest tensor rank, and order of truncation, this yields a finite but systematically constructed search space.

Once the symbolic term libraries have been constructed, each term in the library is evaluated in the weak form over many space-time subdomains across the input data set, which reduces sensitivity to numerical differentiation error \citep{gurevich2019robust}. For each irreducible representation, this defines an overdetermined linear system of the form
\begin{eqnarray}\label{eq:SPIDER_H}
    c_1F_1+\cdots+ c_NF_N = 0,
\end{eqnarray}
where $F_n\in\mathcal{L}_m$ are symbolic terms and $c_n$ are numerical coefficients. Homogeneous sparse regression is then used to identify parsimonious relations of the form \eqref{eq:SPIDER_H} with low residuals and only a small number of nonzero coefficients. 

Any such relation explicitly respects the symmetries of the system. Under a symmetry transformation corresponding to a group element $g$, equation \eqref{eq:SPIDER_H} transforms as
\begin{eqnarray}\label{eq:SPIDER_T}
    c_1\mathcal{S}_m(g)F_1+\cdots+ c_N\mathcal{S}_m(g)F_N = 0,
\end{eqnarray}
where $\mathcal{S}_m(g)$ is a transformation operator specific to the representation. Equivariance means that equation \eqref{eq:SPIDER_T} can be transformed back to the form of equation \eqref{eq:SPIDER_H} by applying the operator $\mathcal{S}_m^{-1}(g)=\mathcal{S}_m(g^{-1})$, so the two equations are mathematically equivalent although not necessarily identical. In particular, for a scalar representation, $\mathcal{S}_m(g)=\mathbb{1}$ for all $g$ and equivariance implies that the form of the equation is invariant under group transformations.

It is possible to include additional terms in the appropriate libraries manually. For instance, in order to find a parameterization of the SGS stress tensor in terms of the specified variables, $\tau_{ij}$ can be included in the rank-2 symmetric trace-free library. In this case, SPIDER can be used to look for a parsimonious relation of the form  
\begin{eqnarray}\label{eq:SPIDER_I}
    \tau=c_1F_1+\cdots+ c_NF_N
\end{eqnarray}
via inhomogeneous sparse regression.
We found both homogeneous and inhomogeneous regression to be valuable in constructing a SGS model.

While the functional form of the governing equations making up a sufficiently general SGS model should not depend on the choice of the data set, the numerical coefficients $c_n$ generally will. In step (4), our objective is to find how these coefficients (i.e., parameters of the coarse-grained description) relate to the parameters of the original problem, i.e., viscosity $\nu$ and filter scale $\Delta$. These relations can be recovered systematically using a combination of scaling and dimensional analysis. For instance, repeating the regression for several data sets covering a range of filter scales, one can determine how every coefficient scales with $\Delta$. One can similarly determine the scaling with $\nu$. Alternatively, the scaling with one (or both) parameters can be deduced using dimensional analysis. Once the scaling has been determined, the results of the regression can be used to find the universal nondimensional numerical prefactors.

A combination of scaling and dimensional analysis can also inform us when our choice of model variables is problematic. For instance, dimensional analysis may predict scaling inconsistent with regression results. Dimensional analysis may also predict unphysical scaling: for instance, no coefficient should diverge in the limit of vanishing $\nu$ or $\Delta$. If found, such inconsistencies provide an indication of incompleteness of the chosen set of model variables or of non-analytic dependencies on invariants. An example of this situation will be discussed below.  

\section{Results}\label{sec:results}
\subsection{Model inference}\label{section:model}

For each of the three flows, F1, F2, and F3, we begin by applying the Gaussian filtering at a particular scale $\Delta$ to fully-resolved DNS solutions and storing the result on a coarse grid with resolution equal to the filter scale (which is equivalent to additionally imposing a sharp spectral filter). We do this for a collection of $\Delta$ ranging from $\ell_i$ down to the DNS grid resolution. 

{\color{black}We next determine whether the SGS stress tensor $\tau_{ij}$ can be represented explicitly in terms of only the resolved variables. Since $\tau_{ij}$ is symmetric but not necessarily traceless, we construct rank-two symmetric trace-free and scalar libraries using only $\bar u_i$ and $\bar p$ as model variables and apply inhomogeneous sparse regression to $\tau_{ij}$. Across different flows, filter scales, and tensor representations, the same dominant term is always identified:
\begin{equation}
    \tau_{ij} \approx c_1 (\nabla_k \bar u_i)(\nabla_k \bar u_j) \equiv \tau_{ij}^{(2)}.
\end{equation}
Occasionally, additional small terms are identified as well, but since we are looking for a universal parameterization that holds generally, we ignore these for the time being. Both dimensional analysis and numerical scaling for data sets with different filter scales show that $c_1= C_1\Delta^2$ with a nondimensional prefactor $C_1$. Across all data sets we consistently find $C_1\approx 1/12$ or, equivalently, $c_1 \approx \Delta^2/12$. Incidentally, this corresponds to the nonlinear/gradient/Clark model \citep{clark1979,leonard1975} which has $c_1=\Delta^2/12$. The same parameterization has been found in a recent data-driven study by \citet{jakhar2024}.

The observation that, for some data sets, regression identifies additional terms suggests that a more accurate parameterization can be constructed. 
To find it, we repeat inhomogeneous sparse regression for the difference $\tau - \tau^{(2)}$, which reliably yields
\begin{align}
    \tau_{ij}-\tau_{ij}^{(2)} \approx  
    c_2 (\nabla_k\nabla_m\bar{u}_i)(\nabla_k\nabla_m\bar{u}_j)\equiv \tau_{ij}^{(4)}.
    \label{eq:ngm4}
\end{align}
Repeating the scaling and dimensional analysis yields $c_2 \approx \Delta^4/288$. The parameterization $\tau = \tau^{(2)} + \tau^{(4)}$ with $c_2=\Delta^4/288$ corresponds to the so-called NGM4 model, which was recently independently discovered by \citet{jakhar2025} and \citet{Choi2025}.

Several comments are now in order. First, the parameterization $\tau=\tau^{(2)} + \tau^{(4)}$ has no adjustable parameters and has been discovered from DNS data with no empirical assumptions whatsoever. Second, the scaling of the two terms with $\Delta$ explains why inhomogeneous regression for $\tau_{ij}$ reliably identifies $\tau^{(2)}$ but only occasionally identifies $\tau^{(4)}$. As $\Delta$ is decreased, $\tau^{(4)}$ becomes negligibly small compared to $\tau^{(2)}$ and is obscured by the inaccuracies associated with both the DNS data and the evaluation of derivatives and quadratures on a coarse grid. The two terms become comparable when the filter scale is chosen equal to an integral scale defined via
\begin{equation}
\ell_i \;=\;\left(
\frac{\int\left\langle \left\lVert \tfrac{1}{12}\,(\nabla_k u_i)(\nabla_k u_j)\right\rVert_F \right\rangle dt}
{\left\langle \int\left\lVert \tfrac{1}{288}\,(\nabla_k\nabla_l u_i)(\nabla_k\nabla_l u_j)\right\rVert_F \right\rangle dt}
\right)^{1/2},
\end{equation}
where $\lVert\cdot\rVert_F$ is the Frobenius norm over tensor indices and $\langle\cdot\rangle$ denotes spatial averaging. 
Defined this way, $\ell_i$ can be thought of as the length scale of the relevant coherent structures, e.g., vortices and vorticity filaments.


As shown in \autoref{appendix: Moment Expansion}, the parameterization \eqref{eq:ngm4} can be obtained by performing the formal moment expansion of $\tau$ in the limit of $\delta\equiv\Delta/\ell_i\ll 1$ and retaining the two leading order terms. In this limit, the term $\tau^{(2)}$ represents the dominant contribution to the SGS stress tensor and $\tau^{(4)}$ represents a correction. To see whether an even more accurate parameterization can be constructed, we repeat inhomogeneous regression for the difference $\tau - \tau^{(2)} - \tau^{(4)}$. In contrast to the first two iterations, however, sparse regression does not reliably identify any parsimonious relations. This provides direct data-driven evidence that the chosen observable set is incomplete and that the remaining residual cannot be represented accurately using the resolved variables $\bar{u}, \bar{p}$ alone. The accuracy of the parameterization therefore cannot be improved further without expanding the set of model variables. Indeed, the formal moment expansion is only valid in the continuum limit and is expected to break down on a discrete grid. We therefore write
\begin{equation}\label{eq:tauR}
 \tau_{ij} = \frac{\Delta^2}{12}(\nabla_k \bar u_i) (\nabla_k \bar u_j) + \frac{\Delta^4}{288}(\nabla_k\nabla_m\bar{u}_i)(\nabla_k\nabla_m\bar{u}_j) + R_{ij}
\end{equation}
and treat $R_{ij}$ as an additional SGS variable describing the unresolved scales. This parameterization is closely related to the LCR decomposition of the SGS stress tensor, where Leonard stress tensor $L$ describes the interaction of resolved scales, cross stress tensor $C$ describes the interaction between resolved and subgrid scales and Reynolds stress tensor $R$ describes the interaction between subgrid scales (see \autoref{appendix: Moment Expansion}). The sum $L+C$ is closely approximated by the first two terms on the right-hand-side of equation \eqref{eq:tauR}, so it is natural to interpret the last term as the Reynolds stress tensor.

The introduction of a new tensor variable requires an additional tensor-valued equation to close the system. We therefore enlarge the set of variables supplied to SPIDER to include $\bar u_i$, $\bar p$, and $R_{ij}$ in constructing the associated rank-two symmetric trace-free and scalar libraries. Performing homogeneous regression, we robustly find an evolution equation of the form 
\begin{equation}\label{eq:r_spider}
    \partial_t R_{ij} = - c_1\bar{u}_k \nabla_k R_{ij}
    + c_2(R_{ik}\nabla_k \bar{u}_j
    + R_{jk}\nabla_k \bar{u}_i)
    + c_3 \nabla^2 R_{ij}
    - c_4\ R_{ij}
\end{equation}
across all data sets. Both irreducible representations yield the same evolution equation with near-identical coefficients, demonstrating consistency across tensor representations.
While reminiscent of Reynolds-stress transport models, equation~\eqref{eq:r_spider} is derived entirely from filtered DNS data without using either phenomenological assumptions or prescribed pressure–strain relations.

The remaining step is to convert the numerical coefficients returned by SPIDER into functional relations. We repeat the scaling analysis over data sets with different $\Delta$ and $\nu$ and analyze whether the inferred numerical coefficients closely follow some dimensionally admissible scaling. 
In particular, $c_1$ and $c_2$ do not scale with either $\Delta$ or $\nu$, are dimensionless, and have values close to unity; therefore, we set $c_1=c_2=1$. Indeed, $c_1=1$ reflects Galilean invariance, which we infer from data rather than assuming it at the outset. The coefficient $c_3$ has the same dimensions as $\nu$ and observed scaling $c_3=C_3\nu$ with $C_3\approx 1$ for all data sets. Consequently, we set $c_3=\nu$.

The coefficient $c_4>0$ has units of inverse time and must scale as $\nu/\Delta^2$ if it depends only on the parameters $\nu$ and $\Delta$. This scaling with $\Delta$ is unphysical. Moreover, the numerical value of $c_4$ is found to scale with neither $\Delta$ nor $\nu$; instead, we consistently find  $c_4=O(\omega_{\rm rms})$. 
The observed scaling suggests that $c_4$ is not a constant but rather a scalar function of the other model variables. Furthermore, $c_4$ cannot be a function of $R_{ij}$ since $R_{ij}$ itself scales with the filter scale as $\Delta^6$. We do not need to consider $\bar{p}$, which is itself related to $\bar{\bf u}$ via the pressure-Poisson equation. Galilean invariance therefore implies that $c_4$ should be a function of spatial derivatives of $\bar{\bf u}$. 
Rotational invariance and dimensional analysis further imply that $c_4$ is a function of invariants of the tensors $\nabla_j\bar{u}_i$, $\Delta\nabla_j\nabla_k\bar{u}_i$, $\Delta^2\nabla_j\nabla_k\nabla_l\bar{u}_i$, etc. The dominant (in terms of the small parameter $\delta$) contribution is a function of invariants of $\bar S$ and $\bar\Omega$, the symmetric and antisymmetric parts of $\nabla\bar{\bf u}$. 
We consider the two lowest-order invariants, $I_1 = 2\bar{S}_{ij}\bar{S}_{ij}$ and $I_2 = 2\bar{\Omega}_{ij}\bar{\Omega}_{ij}$  \citep{pope_more_1975}.

The Buckingham Pi theorem then implies that any admissible relation for $c_4$ can be written as $c_4=\sqrt{I_1}f(I_1/I_2)$, where $f$ is a dimensionless scalar function. We therefore augment the set of variables supplied to SPIDER with the scalar invariants $\sqrt{I_1}$, $\sqrt{I_2}$, and $\sqrt{I_1+I_2}=\sqrt{I_1}\sqrt{1+I_2/I_1}$ and repeat sparse regression. Both homogeneous and inhomogeneous regression robustly identifies
\begin{equation}\label{eq:c4}
    c_4 = \alpha \sqrt{I_1 + I_2} \equiv \alpha I 
\end{equation}
with a nondimensional prefactor $\alpha$ as the best (although far from perfect) approximation.
The identified form of the relaxation term $-c_4R_{ij}$ admits a simple physical interpretation. Since
$I^2= I_1+I_2  =\nabla_i \bar u_j \nabla_i \bar u_j$,
we have $c_4 \sim \|\nabla \bar{\mathbf u}\|_2$. Therefore, $R_{ij}$ relaxes on the time scale $\tau_R=c_4^{-1}$ set by the inverse of the local strain and vorticity of the resolved flow.

The inferred values of $\alpha$ vary moderately across the training data sets, taking values in the range $0.1$--$0.25$. We interpret this range as an estimate of the model-form uncertainty associated with representing the unresolved contribution to the $R_{ij}$ dynamics by the single leading-order invariant $I=\sqrt{I_1+I_2}$. In other words, the functional dependence $c_4\propto I$ is robustly identified, but the nondimensional prefactor is not universal to the same degree as, say, the coefficients $c_1$ and $c_2$. This variation reflects differences between data sets rather than regression uncertainty, and indicates that equation \eqref{eq:c4} should be interpreted as an approximation based on the chosen set of variables (i.e., $\bar{p}$, $\bar{u}_i$, $R_{ij}$, $I_1$ and $I_2$). We use a fixed value of $\alpha=1/4$ below unless explicitly noted otherwise.


Substituting the inferred values of the coefficients into equation \eqref{eq:r_spider}, we finally obtain an evolution equation
\begin{equation}\label{eq:R}
    \partial_t R_{ij} = -\bar{u}_k \nabla_k R_{ij}
    + R_{ik}\nabla_k \bar{u}_j
    + R_{jk}\nabla_k \bar{u}_i
    + \nu \nabla^2 R_{ij}
    - \alpha I\ R_{ij}.
\end{equation}
We will refer to the closed system of equations \eqref{eq:tauR} and \eqref{eq:R} as the NGMR model. }

\subsection{A priori Benchmarks}

In order to quantify how accurate the resulting data-driven model is, let us define a magnitude-aware cross-correlation between two tensor fields $A$ and $B$ by
\begin{equation}
  {\rm CC}(A,B)
  = \frac{\int\langle A_{ij\cdots} B_{ij\cdots}\rangle dt}
         {\max\!\left(\int\langle A_{ij\cdots}A_{ij\cdots}\rangle dt,
                      \int\langle B_{ij\cdots}
                      B_{ij\cdots}\rangle dt\right)}.
  \label{eq:C_AB}
\end{equation}
While remaining bounded by unity, the correlation defined in this way penalizes the magnitude mismatch, providing a stricter measure of similarity than the standard Pearson correlation.
We can now define the accuracy of an LES model in reproducing a tensor field $\phi$ computed from (filtered) DNS data as
\begin{equation}
  \mathcal{A}(\phi)
  = {\rm CC}(\phi^{\rm DNS},\phi^{\rm LES})
  \label{eq:accuracy}
\end{equation}
with $\mathcal{A}(\phi)=1$ indicating an exact match. 

\begin{figure*}
    \centering
    
    \begin{subfigure}{0.32\textwidth}
    \includegraphics[width=\textwidth]{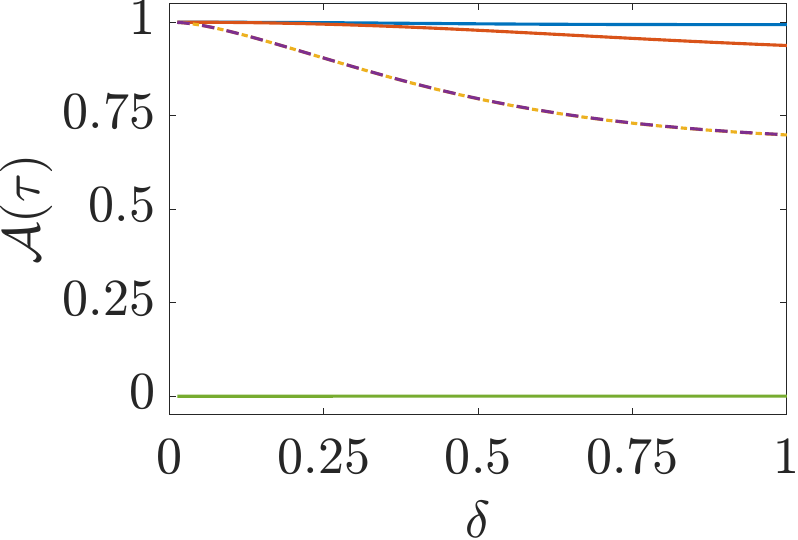}
    \end{subfigure}
    \hspace{1mm}
    \begin{subfigure}{0.32\textwidth}    \includegraphics[width=\textwidth]{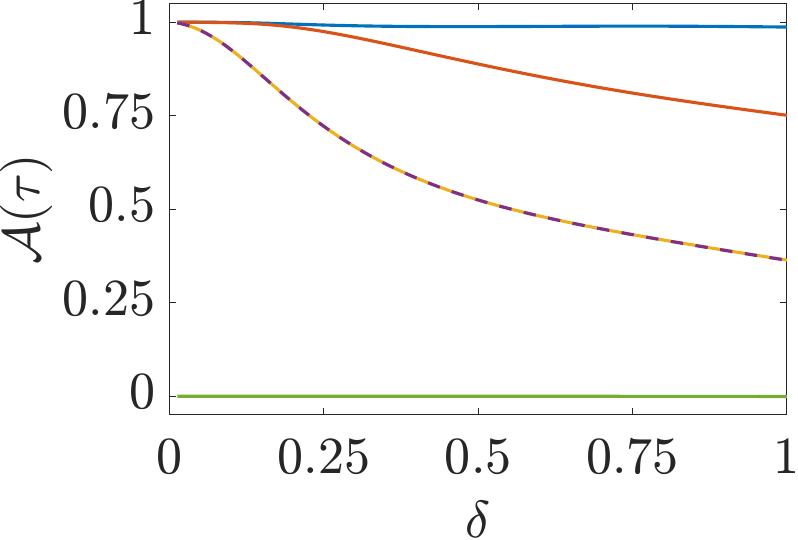}
    \end{subfigure}
    \hspace{1mm}
    \begin{subfigure}{0.32\textwidth}
    \includegraphics[width=\textwidth]{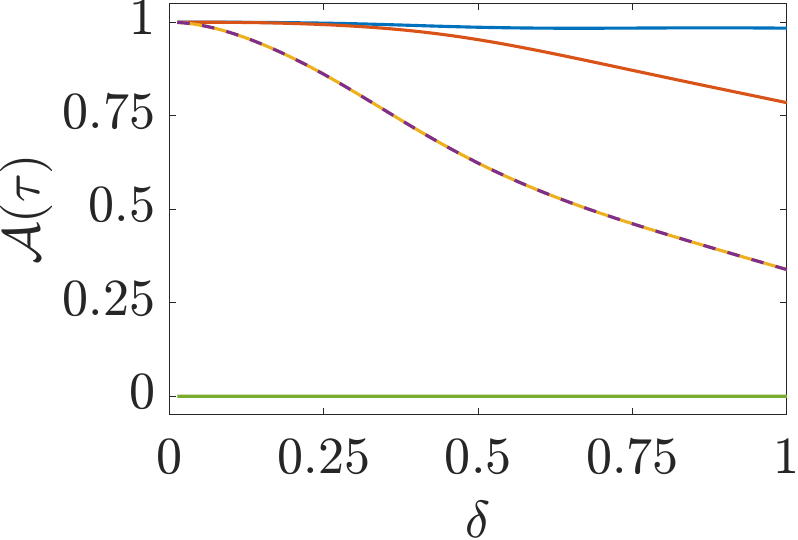}
    \end{subfigure}
    
    \begin{subfigure}{0.32\textwidth}\includegraphics[width=\textwidth]{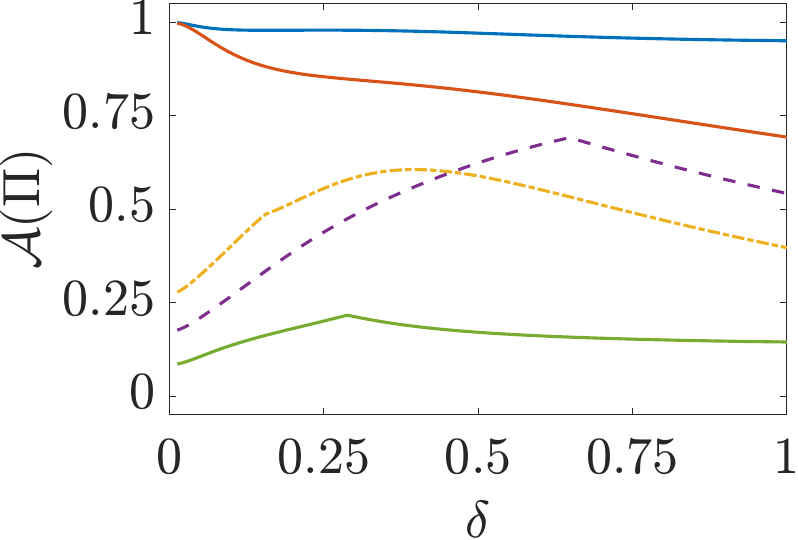}
    \end{subfigure}
    \hspace{1mm}
    \begin{subfigure}{0.32\textwidth}\includegraphics[width=\textwidth]{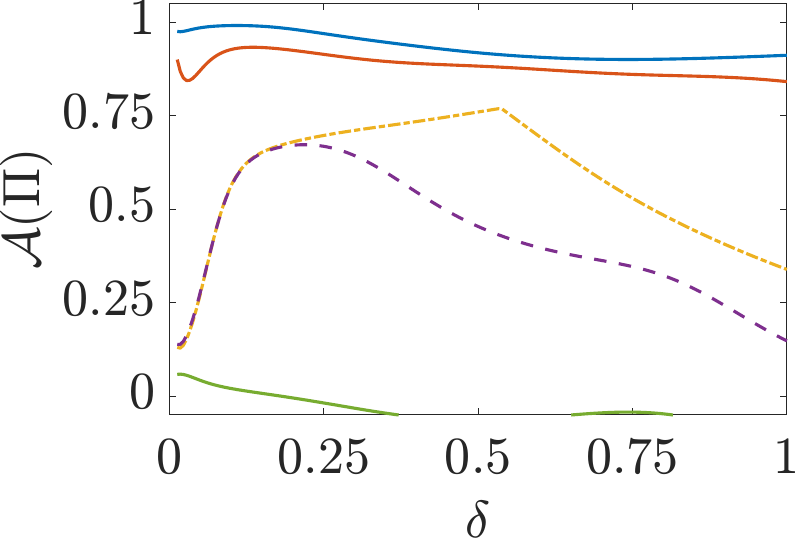}
    \end{subfigure}
    \hspace{1mm}
    \begin{subfigure}{0.32\textwidth}\includegraphics[width=\textwidth]{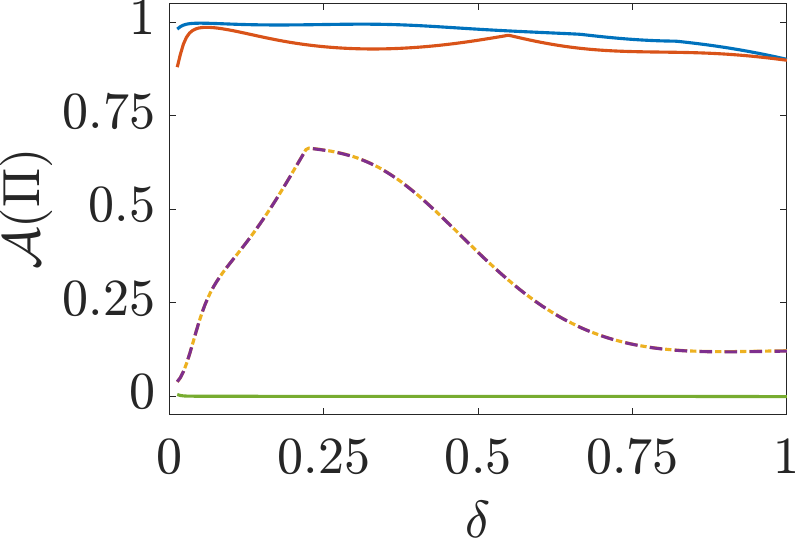}
    \end{subfigure}

    \hspace{3mm}
    \begin{subfigure}{0.3\textwidth}\includegraphics[width=\textwidth]{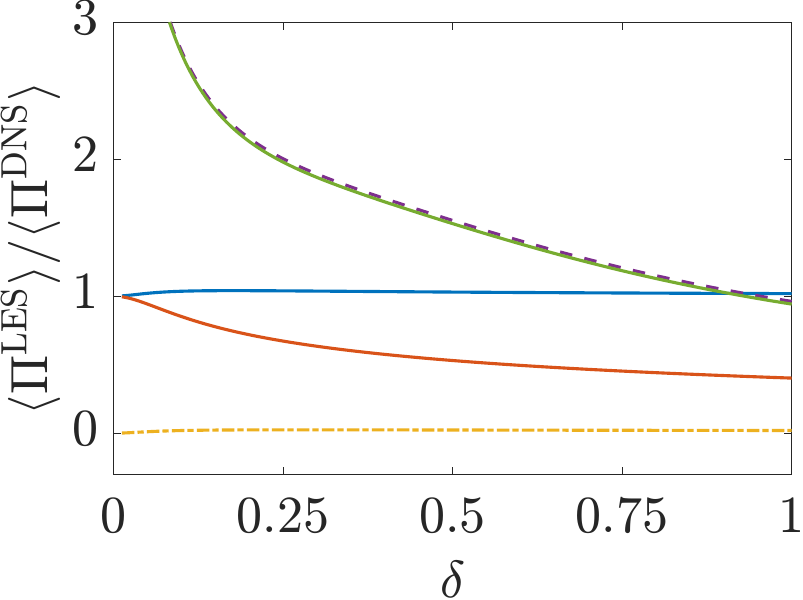}
    \caption{}
    \end{subfigure}
    \hspace{3mm}
    \begin{subfigure}{0.3\textwidth}\includegraphics[width=\textwidth]{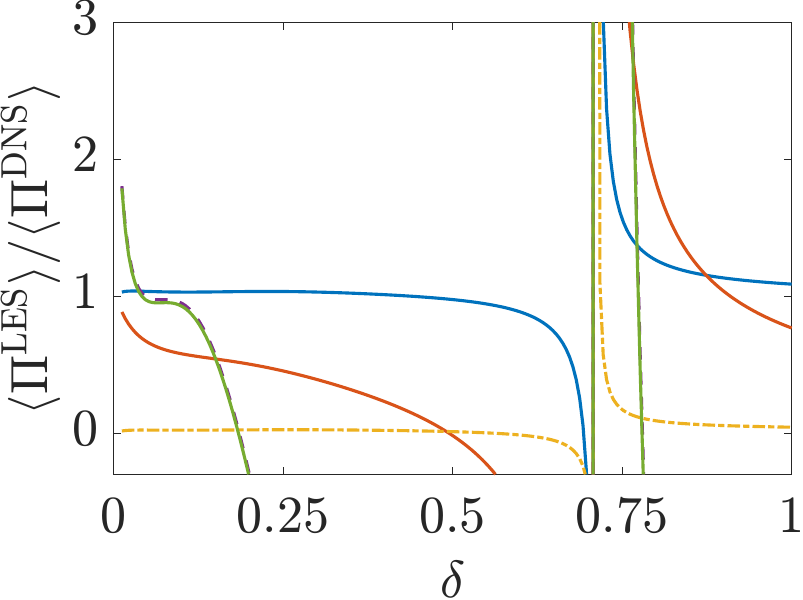}
    \caption{}
    \end{subfigure}
    \hspace{4mm}
    \begin{subfigure}{0.3\textwidth}\includegraphics[width=\textwidth]{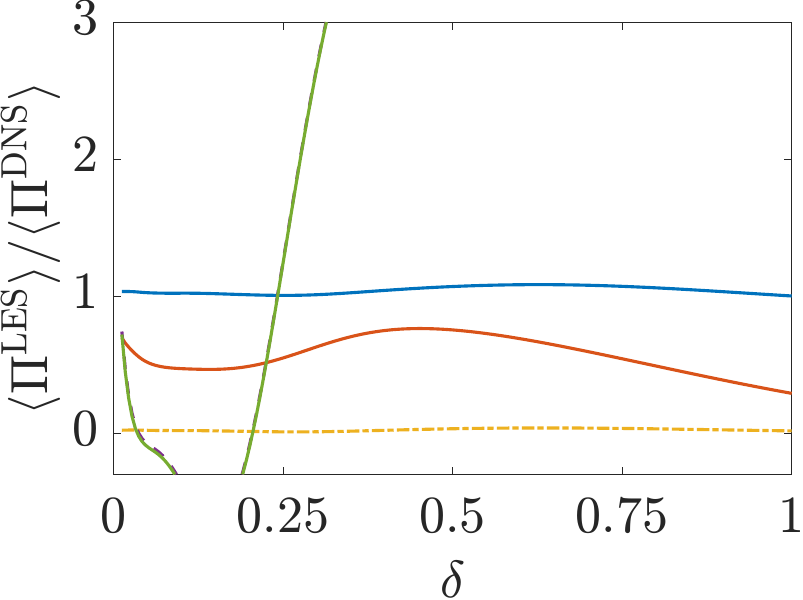}
    \caption{}
    \end{subfigure}
    
    \caption{{\em A priori} benchmarks for the flows F1 (a), F2 (b), and F3 (c) with $\nu = 10^{-5}$: the accuracy of the SGS stress tensor $\mathcal{A}(\tau)$, local energy flux $\mathcal{A}(\Pi)$, and mean energy flux normalized by the DNS value $\langle\Pi^{\rm LES}\rangle/\langle\Pi^{\rm DNS}\rangle$. Five LES models are compared: NGMR (blue), NGM4 (orange), similarity (yellow), unclipped dynamic mixed (purple), and unclipped dynamic Smagorinsky (green), each averaged in time over the data set. 
    }
    \label{fig:apri_corr}
\end{figure*}

To assess {{\em a priori}} performance of our data-driven LES model, we calculate the accuracy of the SGS tensor $\mathcal A(\tau)$, the local energy fluxes $\mathcal A(\Pi)$, where $\Pi = -\tau_{ij}\,\bar S_{ij}$,
and the ratio of the net energy fluxes $\langle \Pi^{\mathrm{LES}}\rangle / \langle \Pi^{\mathrm{DNS}} \rangle$. These quantities are all computed using filtered DNS data. The results are shown in \autoref{fig:apri_corr} as a function of the (nondimensional) filter scale $\delta$ for each of the three canonical flows F1, F2, and F3. For comparison, we also included the predictions of several other popular LES models: the similarity model, dynamic mixed model, dynamic Smagorinsky model, and NGM4; these are described in Appendix \ref{Appendix: Models} and do not include any modifications required for numerical stability. Unsurprisingly, dynamic Smagorinsky fails to predict the SGS stress tensor quantitatively for any $\delta$. The similarity and dynamic mixed model yield reasonable predictions only for $\delta\lesssim 0.1$. NGM4 improves significantly on these LES models but still trails NGMR, which yields near-perfect predictions across all filter scales considered, all the way up to $\Delta=\ell_i$ ($\delta=1$). 

The weakness of the phenomenological LES models becomes even more apparent when one considers the local energy flux $\Pi$. No phenomenological models can predict $\Pi$ even for very small values of $\delta$. 
On the other hand, NGM4 is found to outperform all of the phenomenological LES models across the entire range filter scales. But the most accurate results are provided by NGMR, which achieves the most accurate results out of all LES models considered; the most noticeable improvement over NGM4 is for the flow F1. Moreover, this is the only model that accurately predicts the flux over the entire range $0\le\delta\le1$. In contrast, NGM4 exhibits odd drops in the accuracy at low $\delta$ for the flows F2 and F3.

Finally, let us consider the predictions for the mean energy flux $\langle\Pi\rangle$. This is a key metric that guides the construction of phenomenological LES models, but despite this, no such model is able to consistently predict the mean flux with even moderate accuracy for any of the 2D flows considered here. NGM4 tends to underpredict the flux by as much as 50\% and yields poor predictions for the flow F3 even at very low $\delta$. NGMR is again found to be the most accurate model by a wide margin, yielding consistent predictions for all flows and filter scales considered.

\begin{figure*}
    \centering
    \begin{subfigure}{0.32\textwidth}
    \includegraphics[width=\textwidth]{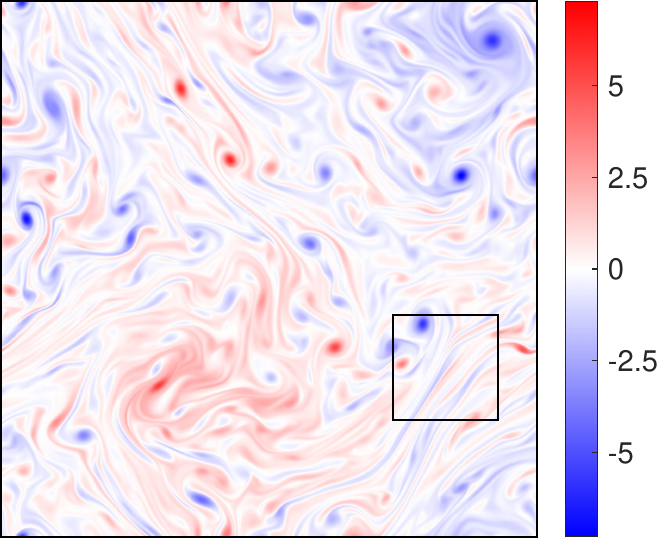}
    \caption{}
    \end{subfigure}
    \hspace{1mm}
    \begin{subfigure}{0.32\textwidth}    
    \includegraphics[width=\textwidth]{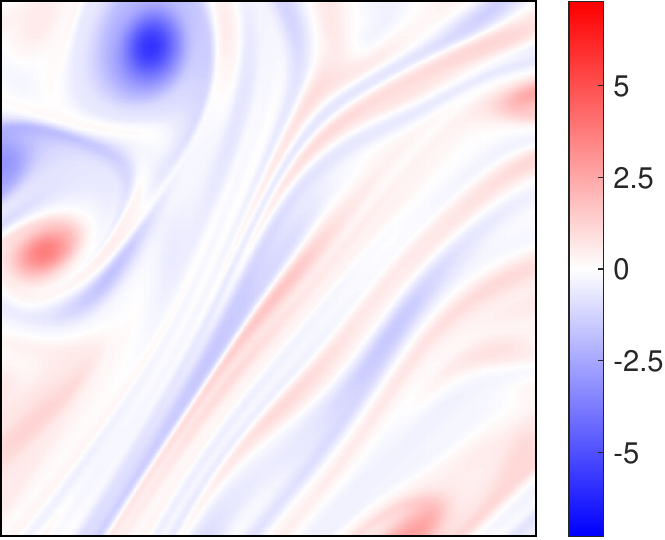}
    \caption{}
    \end{subfigure}
        
    \begin{subfigure}{0.32\textwidth}
    \includegraphics[width=\textwidth]{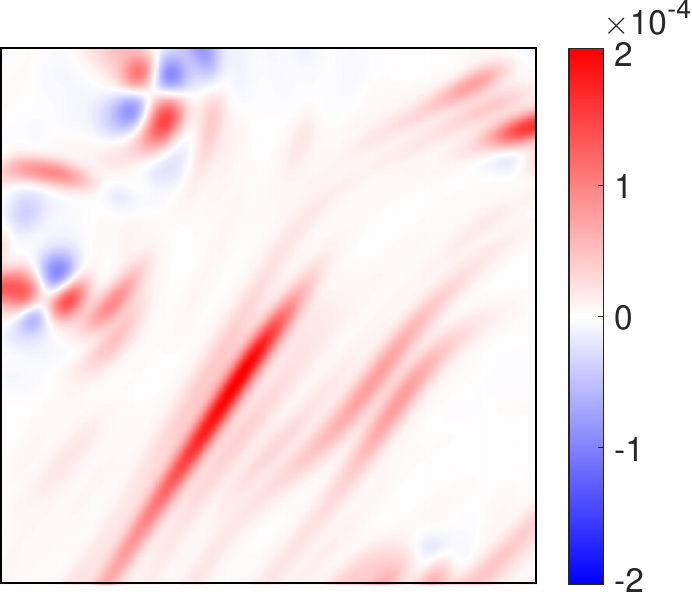}
    \caption{}
    \end{subfigure}
    \hspace{1mm}
    \begin{subfigure}{0.32\textwidth}    
    \includegraphics[width=\textwidth]{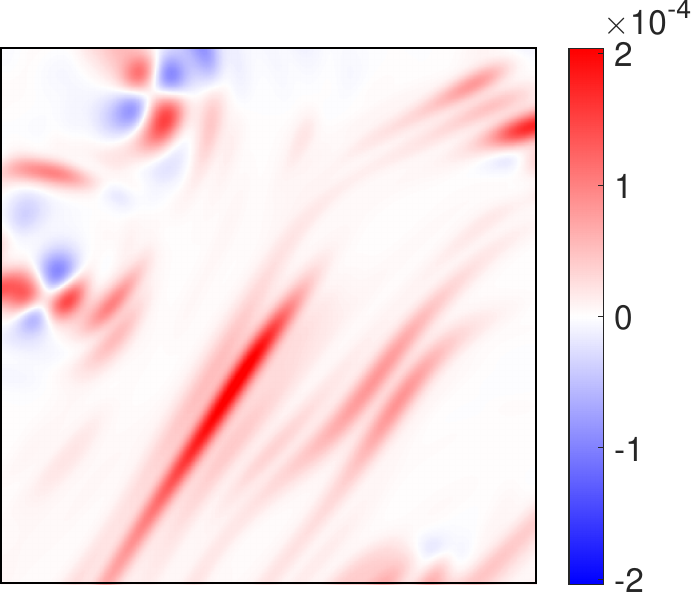}
    \caption{}
    \end{subfigure}
    \hspace{1mm}
    \begin{subfigure}{0.32\textwidth}
    \includegraphics[width=\textwidth]{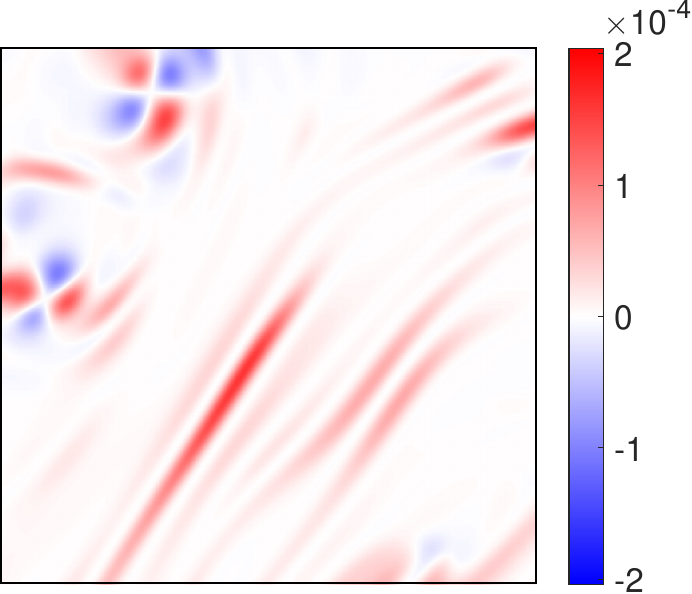}
    \caption{}
    \end{subfigure}
    
    \begin{subfigure}{0.32\textwidth}\includegraphics[width=\textwidth]{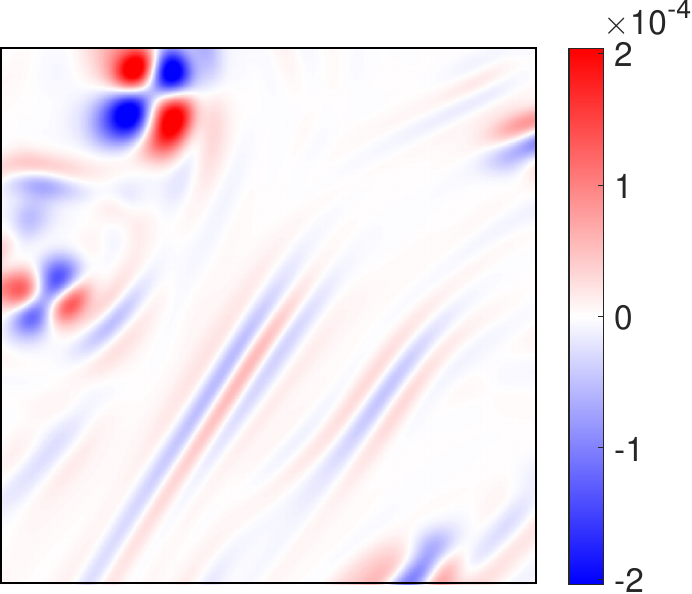}
    \caption{}
    \end{subfigure}
    \hspace{1mm}
    \begin{subfigure}{0.32\textwidth}\includegraphics[width=\textwidth]{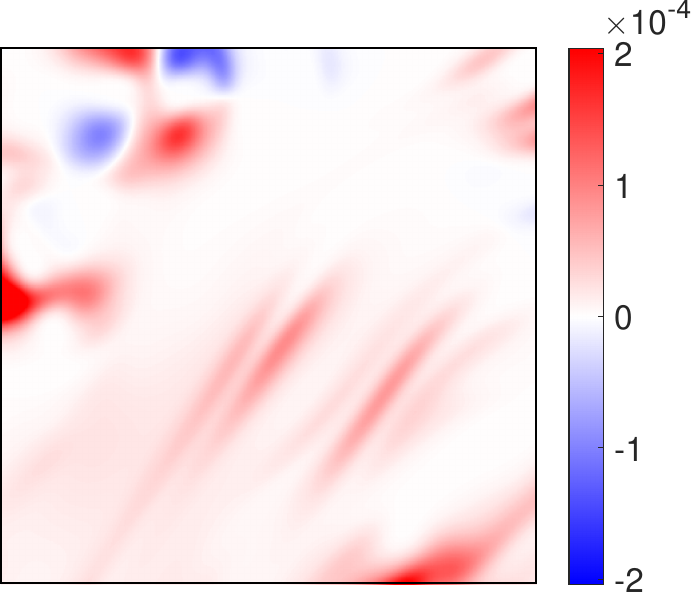}
    \caption{}
    \end{subfigure}
    \hspace{1mm}
    \begin{subfigure}{0.32\textwidth}\includegraphics[width=\textwidth]{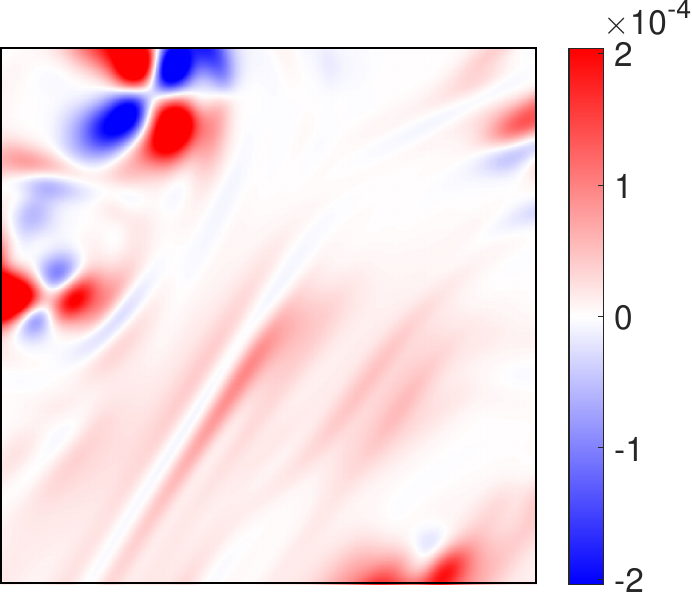}
    \caption{}
    \end{subfigure}
    
    \caption{{{\em A priori}} energy flux for $\delta = 0.3$ for the state shown in panel (a). The black box delineates the region shown in the remaining panels. Panel (b) shows the details of the vorticity field. The remaining panels show the flux corresponding to (c) FDNS, (d) NGMR, (e) NGM4, (f) similarity model, (g) dynamic Smagorinsky, and (h) dynamic mixed model. The flux is shown without backscatter clipping.}
    \label{fig:apriori_fluxes}
\end{figure*}

To provide additional illustration of how the five LES models compare, Figure \ref{fig:apriori_fluxes} shows the spatial structure of the energy flux $\Pi$ predicted by the models compared to DNS in a region of the flow F1 denoted by a square box in panel (a). 
\textcolor{black}{As expected, NGMR reproduces the spatial structure and magnitude of the energy flux much more accurately than the other models considered.} NGM4 nearly matches DNS in the vorticity-dominated region (top left corner) but fails to accurately capture the width of the filamentary structures in the strain-dominated region. The phenomenological models all struggle to reproduce the spatial structure of the energy flux, which explains the low accuracy shown in the middle row of Figure \ref{fig:apri_corr}. The pinwheel structures in the top left corner describe ``breathing modes'' associated with the rotation of elliptic vortices; these co-rotate with the respective vortices. The filamentary structures in the rest of the boxed region represent vortex thinning, the key mechanism for the direct cascade in the extensional regions of the flow. The sign of the flux in this entire region is controlled by the orientation of the vortex filaments (representing small scales) relative to the extensional direction of the large-scale flow \citep{reynoso2024} and here is expected to be the same (positive). 

\begin{figure}
    \centering
    \includegraphics[width=0.35\linewidth]{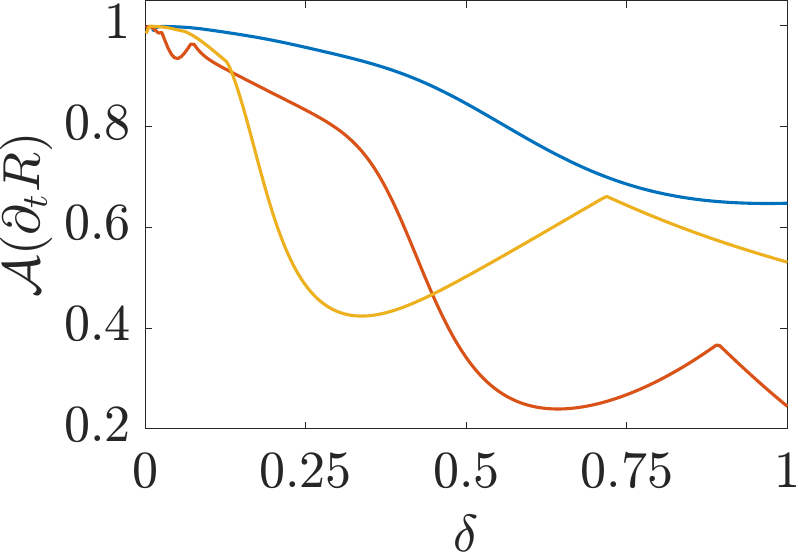}
    \caption{Correlation for the $R$ evolution equation. The data sets correspond to the flows F3
    (blue), 
    F2 
    (orange), and 
    F1 
    (yellow). 
    } 
    \label{fig:r_corr}
\end{figure}

Unlike the rest of LES models, NGMR uses an additional variable to describe the subgrid scales and an associated evolution equation \eqref{eq:R}. The accuracy of this equation is quantified in terms of $\mathcal A(\partial_t R)$. Here, $\partial_t R_{ij}^{\mathrm{DNS}}$ is computed from snapshots of the fully resolved numerical solution using second-order temporal finite differences, while $\partial_t R_{ij}^{\mathrm{LES}}$ is computed by evaluating the right-hand-side of equation \eqref{eq:R} using pseudospectral spatial discretization. The results for a range of filter scales are shown in \autoref{fig:r_corr}. We see that the inferred evolution equation holds to very high accuracy as long as the filter scale is small compared to the size of coherent structures (i.e., $\delta$ is small relative to unity). 
 
\subsection{Stability and Regularization}\label{section: Instabilities and Regularization}
\begin{figure*}
    \centering        
    \begin{subfigure}{0.32\textwidth}
    \includegraphics[width=\textwidth]{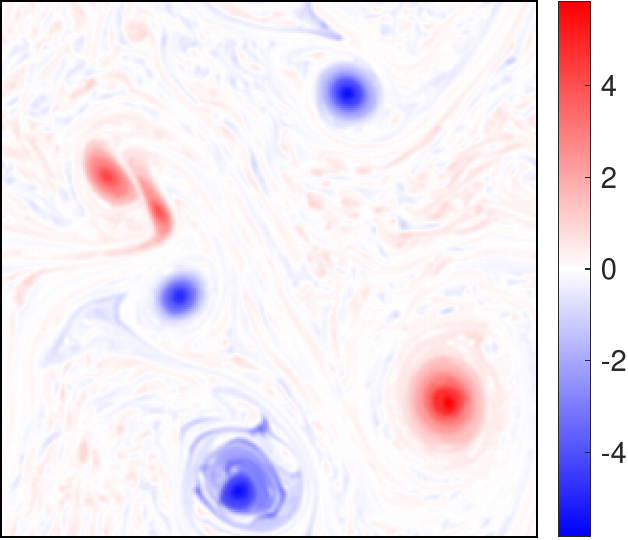}
    \caption{}
    \end{subfigure}
    \hspace{1mm}
    \begin{subfigure}{0.32\textwidth}    
    \includegraphics[width=\textwidth]{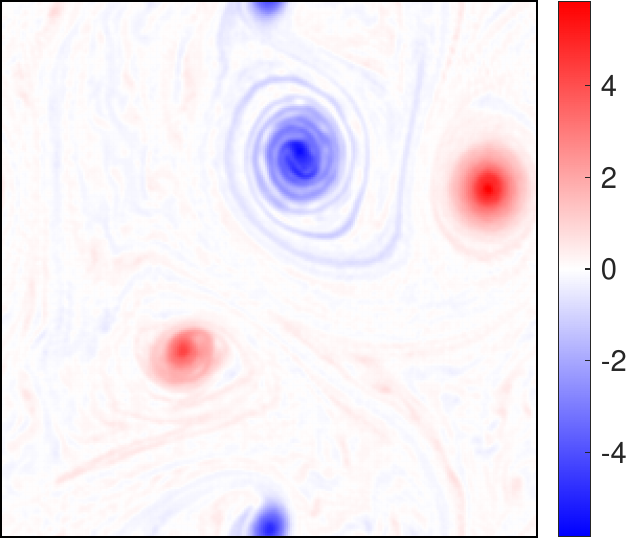}
    \caption{}
    \end{subfigure}
    \hspace{1mm}
    \begin{subfigure}{0.32\textwidth}
    \includegraphics[width=\textwidth]{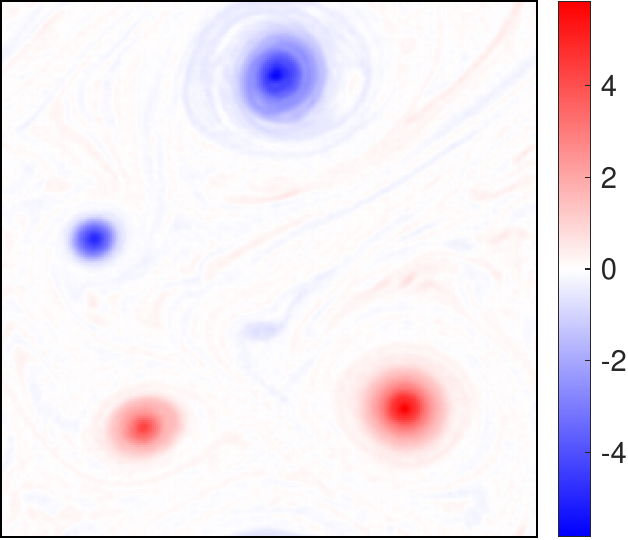}
    \caption{}
    \end{subfigure}
    
    \begin{subfigure}{0.32\textwidth}\includegraphics[width=\textwidth]{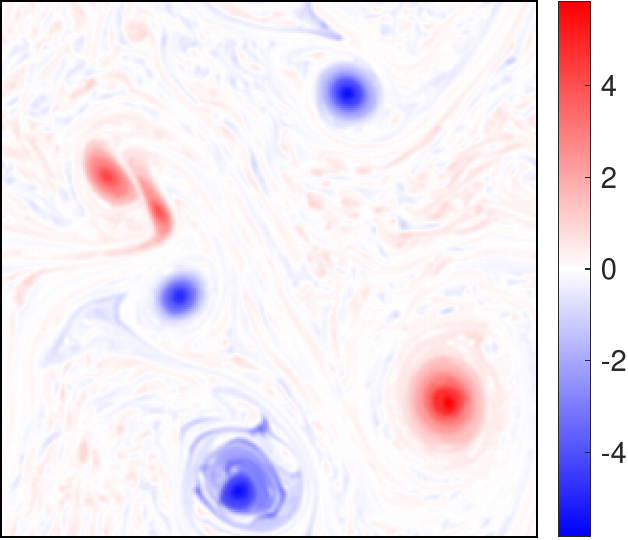}
    \caption{}
    \end{subfigure}
    \hspace{1mm}
    \begin{subfigure}{0.32\textwidth}\includegraphics[width=\textwidth]{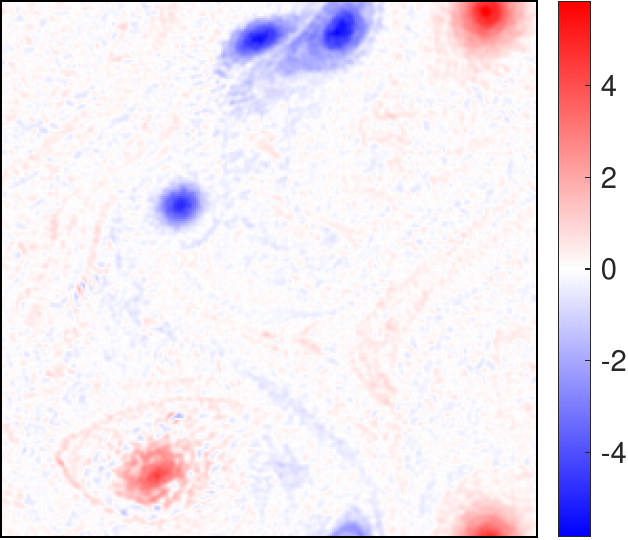}
    \caption{}
    \end{subfigure}
    \hspace{1mm}
    \begin{subfigure}{0.32\textwidth}\includegraphics[width=\textwidth]{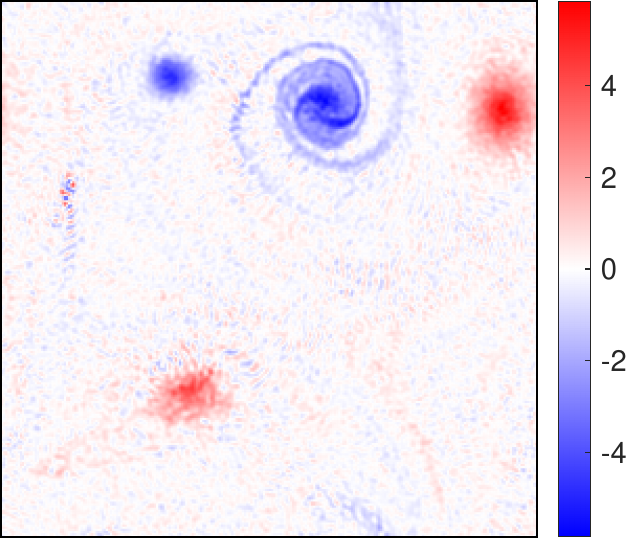}
    \caption{}
    \end{subfigure}
    
    \caption{Snapshots of numerical solutions of NGMR for freely decaying  turbulence at $\nu = 10^{-7}$ on a $256\times256$ grid. The vorticity field computed with (a-c) and without (d-f) the regularizing terms \eqref{reg_1} and \eqref{reg_2} is shown at $t=0$ (a,d), $t=20T_e$ (b,e), and $t=60T_e$ (c,f). }
    \label{fig:reg_vs_noreg}
\end{figure*}

{\color{black}We found that the NGMR model offers improved performance on a variety of {\em a priori} benchmarks. However, using it in an online setting requires a couple of modifications. Such modifications are common in LES models and are typically required to keep numerical solutions stable. For example, it is common to perform various averaging procedures, or outright clip (set to zero) the backscatter contributions, for the dynamic mixed model and dynamic Smagorinsky model \citep{meneveau_scale-invariance_2000}.  
In the case of NGMR, two issues must be addressed. The first one is the linearity of equation~\eqref{eq:R} in $R$ which gives rise to the scaling symmetry $R\to \lambda R$, where $\lambda$ is an arbitrary constant, that is not present in the original, nonlinear system \eqref{eq:gov_eq}.} \textcolor{black}{Scaling invariance implies that asymptotic solutions for $R$ depend on the initial conditions and, furthermore, can vanish or blow up as the flow evolves. }
To address this issue, we separate the Reynolds stress tensor into its traceless part $\mathcal{R}_{ij}$ and the trace ${k}$ via $R_{ij} = \mathcal{R}_{ij} + \tfrac{1}{2}{k}\delta_{ij}$, splitting equation \eqref{eq:R} into two coupled evolution equations
\begin{gather} \label{eq:traceless_r_evo}
    \partial_t\mathcal{R}_{ij} + \bar{u}_l \nabla_l \mathcal{R}_{ij} = \mathcal{R}_{il}\nabla_l \bar{u}_j + \mathcal{R}_{jl}\nabla_l \bar{u}_i - \bar{S}_{lm}\mathcal{R}_{lm}\delta_{ij} + \nu\nabla^2 \mathcal{R}_{ij} - \alpha I\mathcal{R}_{ij} + {k} \bar{S}_{ij},
  \\ \label{eq:trace_r_evo}
  \partial_t{k} + \bar{u}_i \nabla_i {k} = \nu\nabla^2 {k} - \alpha I {k} + 2\bar{S}_{ij}\mathcal{R}_{ij},
\end{gather}
which belong to two different irreducible representations of the rotation group. 

{\color{black}This decomposition formally breaks the linearity of equation \eqref{eq:traceless_r_evo} due to the presence of the term $k \bar S_{ij}$. However, the system is equivalent to equation \eqref{eq:R} and also has a scaling symmetry $\mathcal{R}\to \lambda \mathcal{R}$ and $k\to \lambda k$. To completely break the scaling invariance, we replace the evolution equation for the turbulent kinetic energy $k$ with the leading order term in the corresponding moment expansion (see Appendix \ref{appendix: Moment Expansion})}
\begin{equation}
    {k} = \frac{\Delta^6}{6912}(\nabla_m\nabla^2\bar{u}_i)(\nabla_m\nabla^2\bar{u}_i).
\end{equation}
The trace $k$ can be absorbed into the modified pressure and does not contribute to the energy flux. Hence, inaccuracies associated with the spatial discretization or truncation of the moment expansion do not directly impact the accuracy of the flux predictions. 
The role of turbulent kinetic energy $k$ is mainly to set the scale for $\mathcal{R}$ as determined by the evolution equation \eqref{eq:traceless_r_evo}; this is discussed further in Section \ref{sec:discussion}. 

\textcolor{black}{The second issue is the sensitivity of high frequencies to discretization.} A pseudospectral implementation of the resulting LES model yields stable evolution but exhibits some accumulation of energy in the highest wavenumbers, as illustrated in \autoref{fig:reg_vs_noreg}. To address this issue, regularization terms are introduced in both the filtered momentum equation and the evolution equation for the Reynolds stress tensor. The functional form of the regularization terms is arbitrary, provided they are dimensionally consistent, transform appropriately under the symmetry group, have a negligible impact on the dynamics of large scales, and suppress the energy accumulation.
{\color{black}For pseudo-spectral discretization, it is convenient, although not necessary, to use hyperviscous regularization, i.e., terms such as $\nabla^{2n}u_i$ for equation \eqref{eq:f_ns} and $\nabla^{2m}\mathcal{R}_{ij}$ for equation \eqref{eq:traceless_r_evo}, where $m>1$ and $n>1$ are integers. 
We choose $n=m=3$ to minimize the impact of regularization terms on low wavenumbers.
The prefactors are then chosen to conform to the constraints listed previously, yielding the following functional form:}
\begin{equation} \label{reg_1}
    D^{(1)}_{i} = \frac{\Delta^5}{20736}(\mathcal{R}_{ij}\mathcal{R}_{ij})^{1/4} \nabla^6 \bar{u}_i \textcolor{black}{\ \sim O(\delta^8),}
\end{equation} 
and 
\begin{equation}\label{reg_2}
    D^{(2)}_{ij} =  \frac{\Delta^6}{1728} I \nabla^6 \mathcal{R}_{ij} \textcolor{black}{\ \sim O(\delta^{12}).}
\end{equation}
We have not examined the sensitivity of the results presented below to specific choices of the regularization parameters, such as the numerical prefactors and the powers $n$ and $m$.

The corresponding regularized 
form of the NGMR model suitable for \textcolor{black}{\textit{pseudo-spectral}} numerical implementation is summarized below:
\begin{subequations}\label{ngmR}
\begin{align} 
\label{eq:f_ns_reg}
    \partial_t \bar{u}_i + \bar{u}_j \nabla_j \bar{u}_i &= - \rho^{-1}\nabla_i \bar{p} + \nu \nabla^2 \bar{u}_i -\gamma \bar u_i+ \bar{f}_i - \nabla_j\tau_{ij} + D^{(1)}_i, 
   \\ 
\label{eq_tau_traceless}
  \tau_{ij} &= 
  \frac{\Delta^2}{12}(\nabla_k \bar u_i) (\nabla_k \bar u_j) + \frac{\Delta^4}{288}(\nabla_k\nabla_m\bar{u}_i)(\nabla_k\nabla_m\bar{u}_j)
  + \mathcal{R}_{ij},
  \\ 
\label{eq: R evo}
  \partial_t\mathcal{R}_{ij} + \bar{u}_l \nabla_l \mathcal{R}_{ij} &= \mathcal{R}_{il}\nabla_l \bar{u}_j + \mathcal{R}_{jl}\nabla_l \bar{u}_i + \nu\nabla^2 \mathcal{R}_{ij} -\bar{S}_{lm}\mathcal{R}_{lm}\delta_{ij} - \alpha I \mathcal{R}_{ij} + k\bar{S}_{ij} + D^{(2)}_{ij},
\end{align}
\end{subequations} 
where the turbulent kinetic energy contribution to the SGS stress tensor has been absorbed into the modified pressure. {\em A posteriori} tests described below show that choosing $\alpha = 1/4$ yields the most stable and accurate predictions. 

While the introduction of an additional evolution equation does increase the computational cost of the  model compared to, say, NGM4, this cost is more than offset by a very substantial increase in the accuracy. Furthermore, we found that the computational cost of NGMR is comparable to that of other LES models, such as the similarity model.

\textcolor{black}{Because NGMR evolves the additional SGS variable $\mathcal{R}_{ij}$, online simulations require an initial condition for this field. While the choice of initial condition has an impact on short time scales, we found that long-term statistics are insensitive to this choice.
The details are provided in \autoref{appendix:R_initialization}.}

\subsection{A posteriori Benchmarks}

In order to quantify the accuracy of the online version of the NGMR model and compare it with existing LES methods, we employ a variety of {\em a posteriori} benchmarks for both freely decaying turbulence and forced turbulence, with and without Rayleigh friction, and over a range of Reynolds numbers and filter scales. {\color{black}The key information is summarized in Table \ref{tab:aposteriori_flow_configurations}.}

\begin{table}\color{black}
\centering
\setlength{\tabcolsep}{4pt}
\renewcommand{\arraystretch}{1.15}
\begin{tabular}{@{}lcccc@{}}
\toprule
& F2 & F2 & F4 & F5 \\
\midrule
Regime
& Free decay
& Free decay
& Direct cascade
& Inverse cascade \\

$\nu$
& $10^{-5}$
& $10^{-7}$
& $10^{-6}$
& $10^{-6}$ \\

$\mathrm{Re}$
& $2\times10^4$
& $2\times10^6$
& $2\times10^6$
& $5\times10^5$ \\

$\ell_i$
& $\ell_d/4$
& $\ell_d/4$
& $\ell_d/28$
& $\ell_d/40$ \\

DNS grid
& $2048^2$
& $8192^2$
& $8192^2$
& $8192^2$ \\

Forcing/friction
& None
& None
& \shortstack{$k_c=4$, $\gamma=0.05$}
& \shortstack{$k_c=32$, $\gamma=10^{-4}$} \\

LES grid
& $128^2$
& $256^2$
& $32^2$--$1024^2$
& $512^2$ \\

$\delta$
& $ 0.09$
& $ 0.05$
& $ 0.08$--$2.6$
& $ 0.23$ \\

Figures
& \shortstack{\ref{fig:aposteriori_evolution},
\ref{fig:aposteriori_1e-5}, \ref{fig: vorticity correlations}(a)}
& \shortstack{\ref{fig:reg_vs_noreg},
\ref{fig:aposteriori_1e-7}, \ref{fig: vorticity correlations}(b)}
& \shortstack{\ref{fig: Forced IC}(a),
\ref{fig: Energy low-freq}, \ref{fig: Vorticity low-freq}}
& \shortstack{\ref{fig: Forced IC}(b),
\ref{fig: spectrum}, \ref{fig: vorticity PDF}, \ref{fig: flux PDF}} \\
\bottomrule
\end{tabular}
\caption{\color{black}Summary of the data sets used for {\em a posteriori} benchmarks.}
\label{tab:aposteriori_flow_configurations}
\end{table}

\subsubsection{Freely Decaying Turbulence}

\begin{figure}
    \centering
    \begin{subfigure}{0.32\textwidth}
        \centering
        \includegraphics[width=\textwidth]{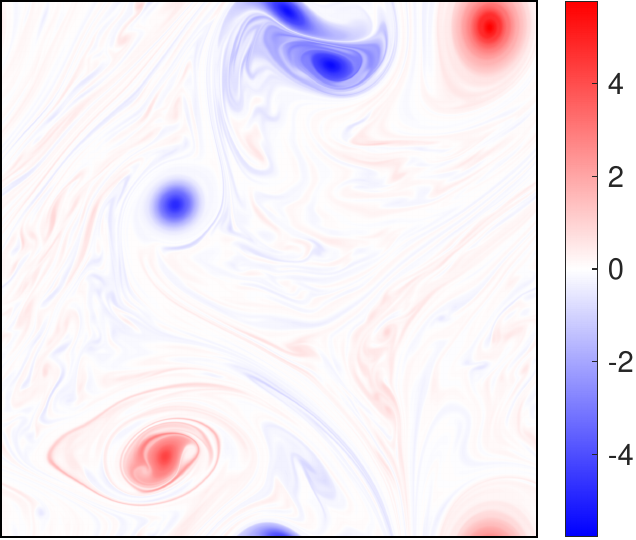}
        \caption{}
    \end{subfigure}
    \hspace{1mm}
    \begin{subfigure}{0.32\textwidth}
        \centering
        \includegraphics[width=\textwidth]{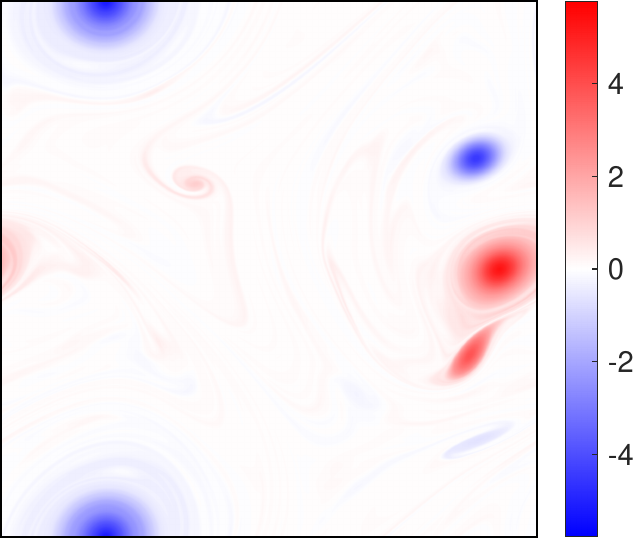}
        \caption{}
    \end{subfigure}
    \hspace{1mm}
    \begin{subfigure}{0.32\textwidth}
        \centering
        \includegraphics[width=\textwidth]{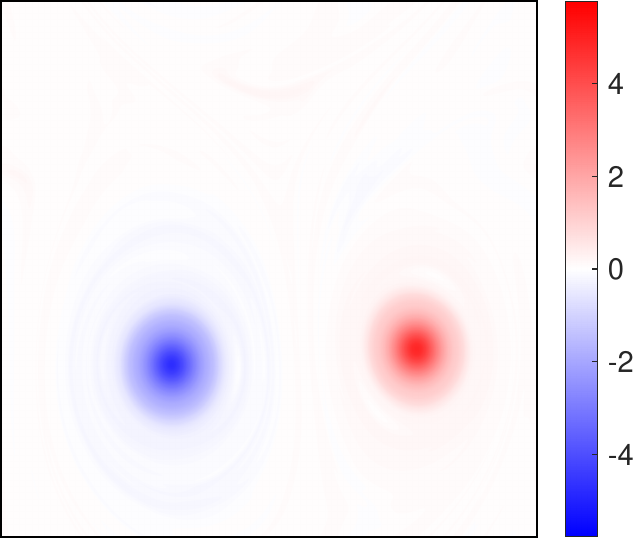}
        \caption{}
    \end{subfigure}

    \caption{Representative snapshots of the $2048\times2048$ DNS solution for the freely decaying flow with $\nu=10^{-5}$ at $t \approx  15T_e$ (a), $t \approx 100T_e$ (b), and $t \approx 250T_e$ (c). 
    }
    \label{fig:aposteriori_evolution}
\end{figure}

We start with the analysis of transient flows represented by freely decaying turbulence without Rayleigh friction. Since turbulence disappears too quickly to collect meaningful statistics for one flow realization, we use an ensemble with 12 qualitatively similar but quantitatively distinct initial conditions. All of these initial conditions have roughly the same eddy turnover time $T_e$, which we will use as a relevant time scale for flow evolution. These initial conditions were constructed using the following procedure: we take the flow F2 shown in \autoref{fig:IC_snapshots}(b) and add a white noise perturbation at an intermediate wavelength of $k=24$ with amplitude of $E(k=24)/5$. The perturbed flow is then evolved
for a time interval of about $30T_e$; this procedure is repeated 12 times to yield distinct initial conditions. {\color{black}In particular, $\mathcal{R}_{ij}$ is initialized using its definition in terms of the fully-resolved velocity field.} 
For each of the 12 initial conditions, DNS is performed for roughly $350T_e$ on a $2048\times 2048$ computational grid for $\nu=10^{-5}$ and $8192\times 8192$ computational grid for $\nu=10^{-7}$, ensuring that the dissipation scale is at least 2$\times$ larger than the grid scale. The LES solutions are also computed for $250T_e$, with computational grids of resolution $128\times 128$ for $\nu=10^{-5}$, or $256\times 256$ for $10^{-7}$. Representative snapshots of the vorticity field for one of the members of the DNS ensemble are shown in \autoref{fig:aposteriori_evolution} and the movies are provided as Supplementary Material. 

\begin{figure*}
    \centering
    
    \begin{subfigure}{0.45\textwidth}
    \includegraphics[width=\textwidth]{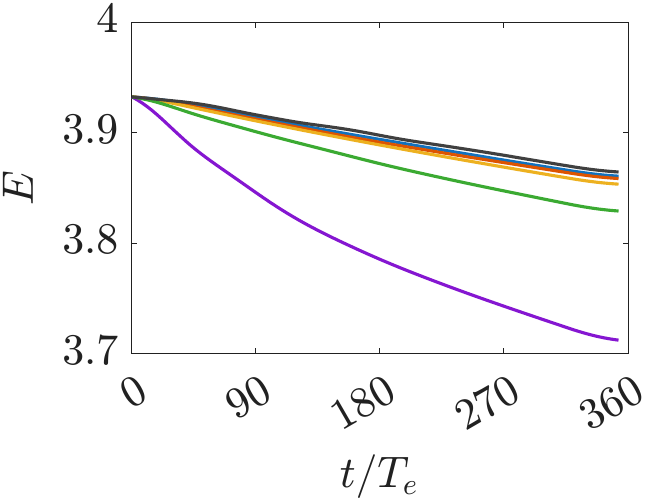}
    \end{subfigure}
    \hspace{5mm}
    \begin{subfigure}{0.45\textwidth}    \includegraphics[width=\textwidth]{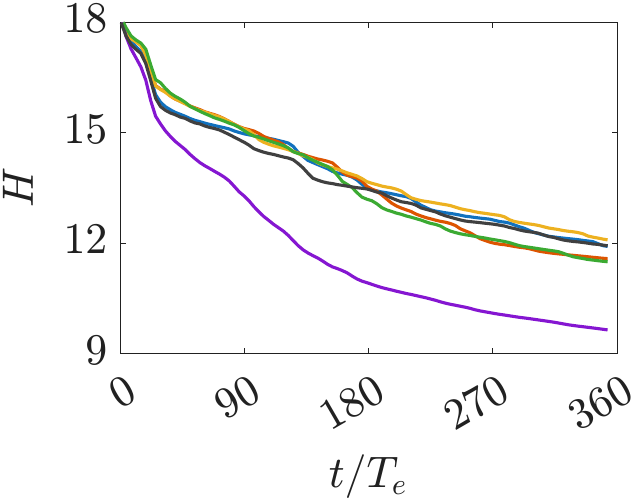}
    \end{subfigure}

    \begin{subfigure}{0.45\textwidth}
    \includegraphics[width=\textwidth]{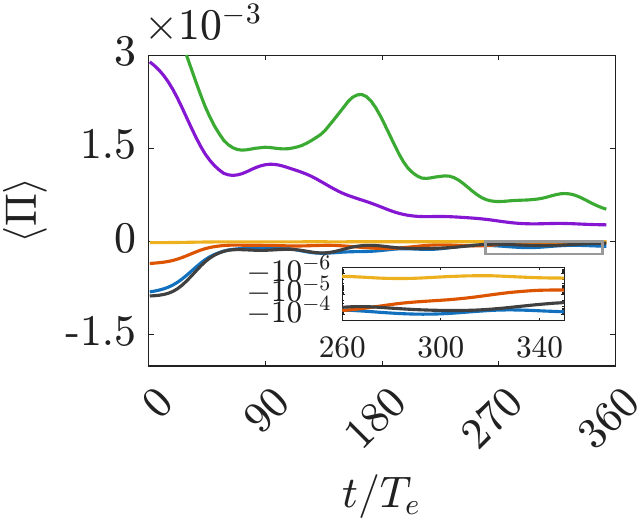}
    \end{subfigure}
    \hspace{5mm}
    \begin{subfigure}{0.45\textwidth}
    \includegraphics[width=\textwidth]{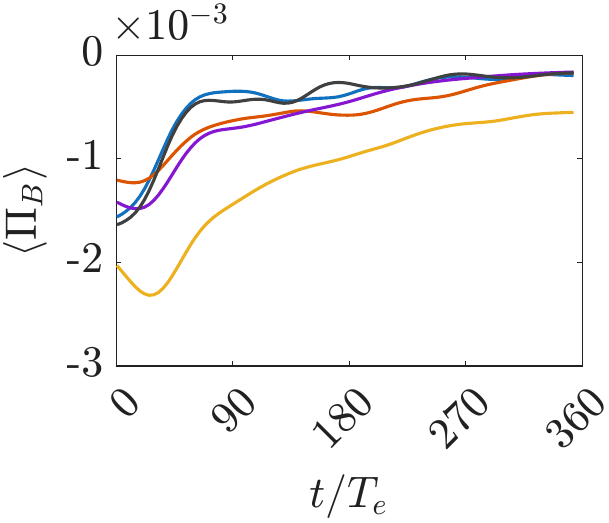}
    \end{subfigure}
    \caption{Ensemble averaged {\em a posteriori} benchmarks for a freely decaying turbulence at $\nu=10^{-5}$: spatial mean of (a) the energy $E$, (b) the enstrophy $H$, (c) the mean energy flux $\langle\Pi_E\rangle$, and (d) the mean backscatter $\langle\Pi_B\rangle$ predicted by LES models and FDNS. Five LES models are compared: NGMR (blue), NGM4 (orange), similarity (yellow), clipped dynamic mixed (purple), and clipped dynamic Smagorinsky (green).  The latter two produce no backscatter due to clipping. }
    \label{fig:aposteriori_1e-5}
\end{figure*}

\begin{figure*}
    \centering
    \begin{subfigure}{0.45\textwidth}
    \includegraphics[width=\textwidth]{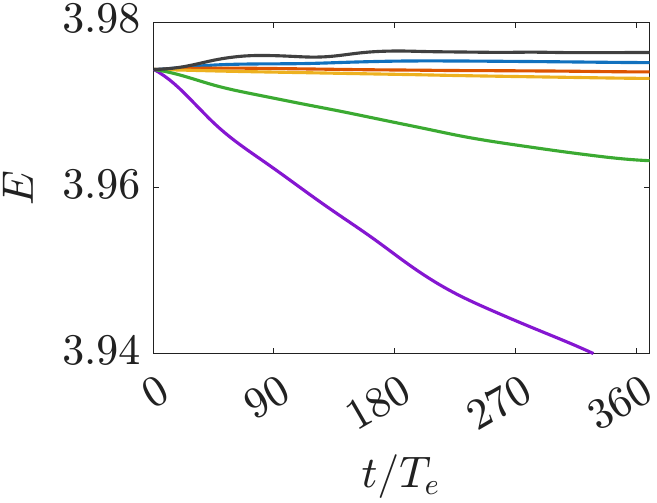}
    \caption{}
    \end{subfigure}
    \hspace{5mm}
    \begin{subfigure}{0.45\textwidth}    
    \includegraphics[width=\textwidth]{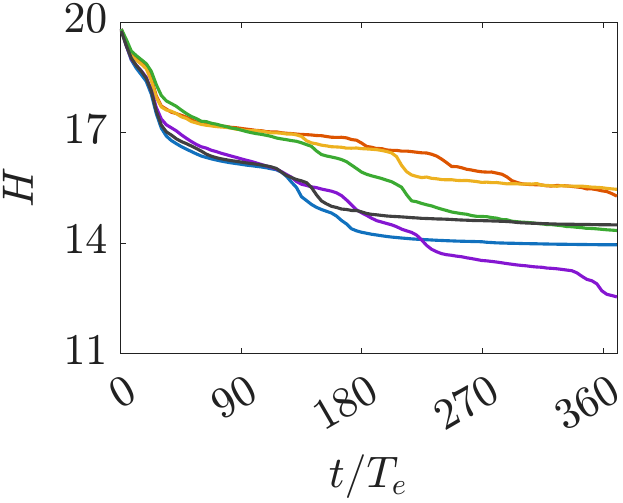}
    \caption{}
    \end{subfigure}

    \begin{subfigure}{0.45\textwidth}
    \includegraphics[width=\textwidth]{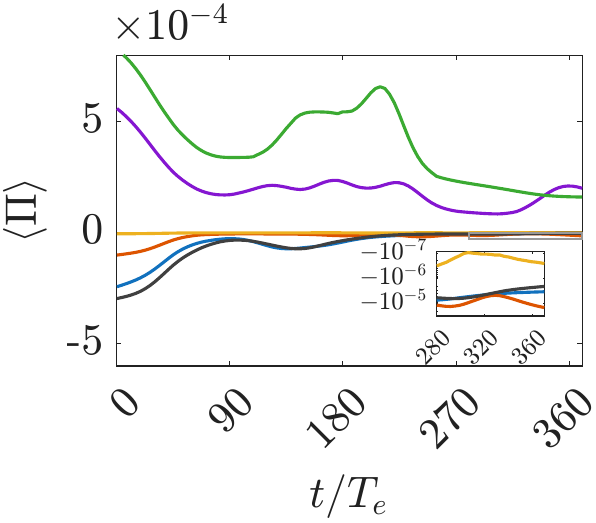}
    \caption{}
    \end{subfigure}
    \hspace{5mm}
    \begin{subfigure}{0.45\textwidth}
    \includegraphics[width=\textwidth]{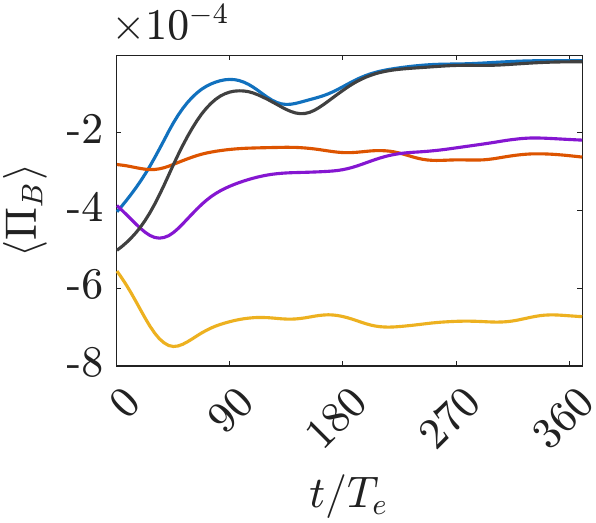}
    \caption{}
    \end{subfigure}

    \caption{Ensemble averaged {\em a posteriori} benchmarks for a freely decaying turbulence. The same quantities are shown as in \autoref{fig:aposteriori_1e-5} but for $\nu=10^{-7}$. {\color{black}The inset in panel (c) corresponds to the region bounded by the gray box.}}
    \label{fig:aposteriori_1e-7}
\end{figure*}

For all numerical solutions (both DNS and SGS), we calculate the spatial means of the energy $E = \tfrac{1}{2}\int_\Omega \bar{\mathbf{u}}\cdot \bar{\mathbf{u}}\, d{\bf x}$, enstrophy $H = \tfrac{1}{2} \int_\Omega \bar{\omega}^2\, d{\bf x}$, energy flux $ \Pi_E = \int_\Omega \Pi \ d{\bf x}$, and backscatter $ \Pi_B = \int_\Omega \Theta(-\Pi) \Pi \, d{\bf x}$, where $\Theta(\cdot)$ is the Heaviside function.
The results are shown in \autoref{fig:aposteriori_1e-5} for $\nu=10^{-5}$ and in \autoref{fig:aposteriori_1e-7} for $\nu=10^{-7}$. 
For both values of viscosity, we find that dynamic Smagorisnky and the mixed model both tend to substantially overestimate the dissipation of energy, which is partially due to the clipping of backscatter required for their stable implementations. NGM4 and the similarity model do better but still underperform considerably compared to NGMR. NGMR also tracks the evolution of enstrophy more accurately than the rest of the models for both values of viscosity, capturing both the overall trend and the quick dips associated with vortex merger events. The dynamic mixed model is again found to be overly dissipative, especially for $\nu=10^{-5}$. The mean energy fluxes -- both net and backscatter -- reveal the most dramatic difference between the models. It is only the NGMR model that consistently matches FDNS. All the other models, including NGM4, consistently yield poor predictions throughout the evolution. Most disturbingly, the errors in predicting the overall energy dissipation are particularly severe for the two models---dynamic Smagorinsky and dynamic mixed---designed specifically to predict this quantity.

\begin{figure}
    \centering
    \begin{subfigure}{0.45\textwidth}
    \includegraphics[width=\textwidth]{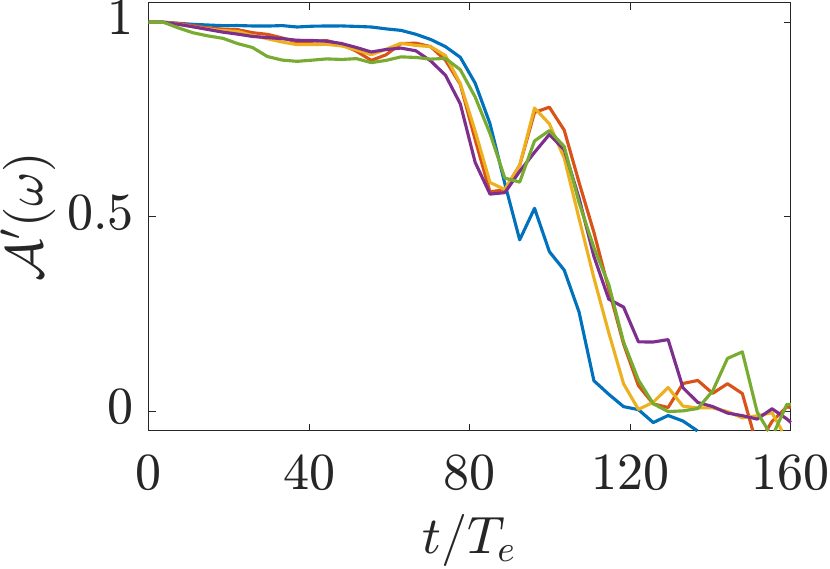}
    \caption{}
    \end{subfigure}
    \hspace{5mm}
    \begin{subfigure}{0.45\textwidth}    \includegraphics[width=\textwidth]{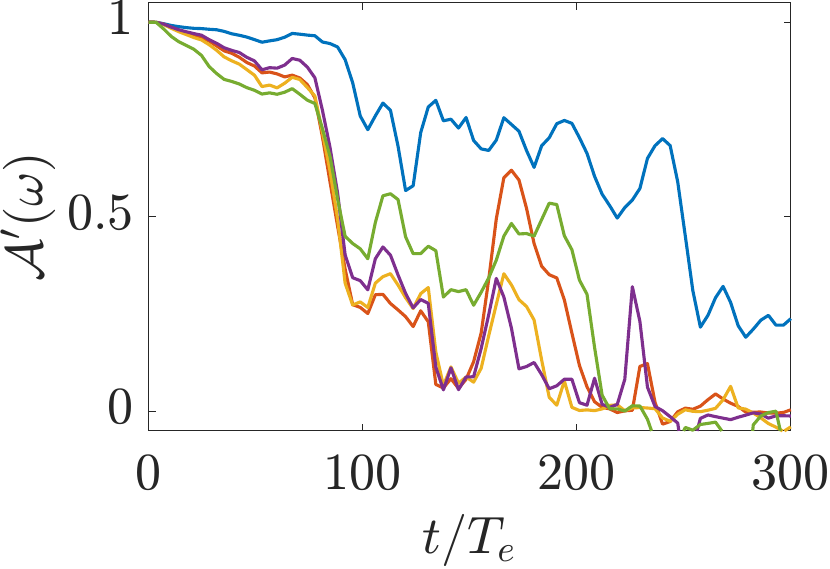}
    \caption{}
    \end{subfigure}
    \caption{Ensemble average of the cross-correlation $\mathcal{A}'({\omega})$ for the vorticity field of freely decaying turbulence at $\nu=10^{-5}$ (a) and $\nu =10^{-7}$ (b). Five models are compared: NGMR (blue), NGM4 (orange), similarity (yellow), clipped dynamic mixed (purple), and clipped dynamic Smagorinsky (green). }
    \label{fig: vorticity correlations}
\end{figure}

To compare short-term forecasting, we also define the accuracy metric 
\begin{equation}
  \mathcal{A}'(\omega)
  = \frac{\langle\bar \omega^{LES} \bar \omega^{DNS}\rangle }
         {\max\! \left(\langle \bar \omega^{LES} \bar \omega^{LES}\rangle,
                      \langle \bar \omega^{DNS} \bar \omega^{DNS}\rangle \right)},
  \label{eq:temp_correlation}
\end{equation}
describing spatial (vs. spatiotemporal, as in the case of equation \eqref{eq:accuracy}) cross-correlation between an FDNS solution and a corresponding LES solution. The correlation is compared for different models in \autoref{fig: vorticity correlations} as a function of time.
For the $\nu=10^{-5}$ case, NGMR yields a solution that remains almost perfectly correlated with DNS for $t/T_e\lesssim 180$ while the correlation for the rest of the models immediately starts to decrease. Interestingly, none of the models can accurately predict the evolution beyond $t/T_e\approx 180$. The results are qualitatively similar for the $\nu=10^{-7}$ case. Not only does NGMR yield the most accurate solutions for $t/T_e\lesssim 200$, the corresponding solution remains strongly correlated with DNS for a period of time as long as $t/T_e\approx 250$ while, for the rest of the models, the predictions decorrelate completely after $t/T_e\approx 80$.

In addition to accuracy benchmarks, we also qualitatively assess the computational efficiency of our closure. Every model was implemented and benchmarked for around $100T_e$ at multiple grid resolutions ranging from $N_{\mathrm{LES}} = 128$ to $1024$. All timing tests were performed using naive implementations while ensuring identical numerical schemes, parallelization settings, and hardware configurations to ensure fair comparison. Overall, NGMR performs comparably with all other models, with NGMR $\approx 2$ times faster than dynamic mixed and dynamic Smagorinsky, and $\approx 0.5$ times slower than NGM4 and similarity, at the same resolutions.

\subsubsection{Forced Turbulence}\label{section:forced}
\begin{figure}
    \centering
    \begin{subfigure}{0.45\textwidth}
        \centering
        \includegraphics[width=\textwidth]{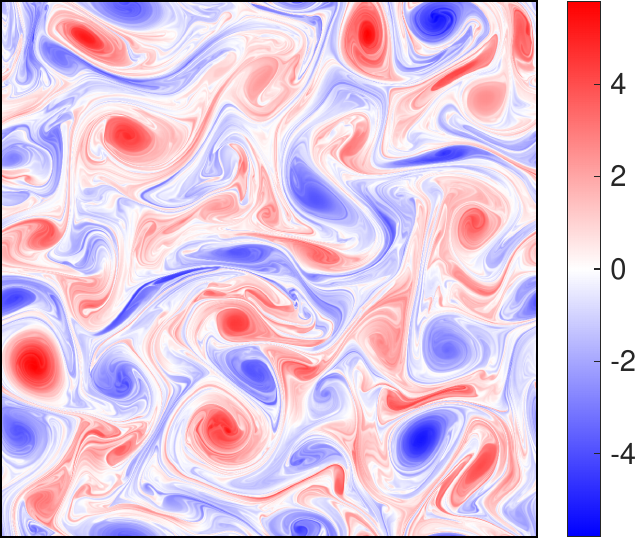}
        \caption{}
    \end{subfigure}
    \hspace{5mm}
    \begin{subfigure}{0.45\textwidth}
        \centering
        \includegraphics[width=\textwidth]{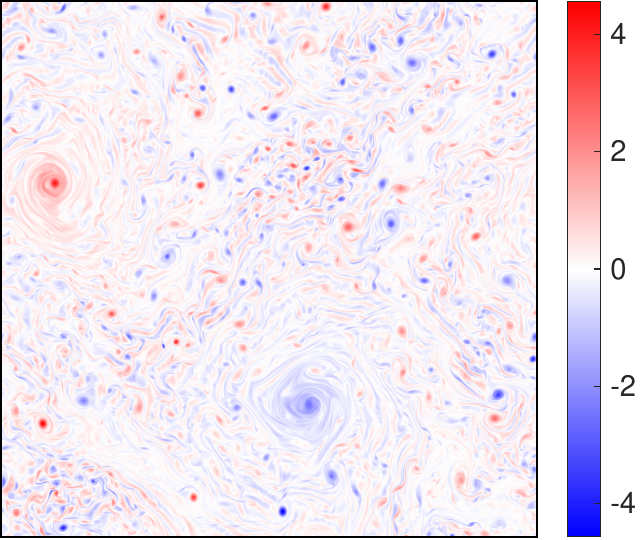}
        \caption{}
    \end{subfigure}

    \caption{Representative snapshots of the two forced flows F4 (a) and F5 (b) with $\ell_i \approx \ell_d/28$ and $\ell_d/40$, respectively.}
    \label{fig: Forced IC}
\end{figure}

Statistically stationary turbulent flows were generated with the forcing profile
\begin{equation}\label{eq:checkerboard}
    \nabla\times \textbf{f} = \frac{1}{2}
    \sin(k_cx)\sin(k_cy)
\end{equation}
and Rayleigh friction, following \citet{jakhar2025}. Note that for this forcing profile, the effective forcing wavelength is given by $k_f=\sqrt{2}k_c$. Representative snapshots of two such flows, labeled F4 and F5,  are shown in Figure \ref{fig: Forced IC}. In both cases, we used a viscosity of $\nu=10^{-6}$. The flow F4 is forced at wavenumber $k_c = 4$ with Rayleigh friction coefficient $\gamma = 0.05$ and corresponds to Re~=~$2 \times 10^6$. The flow F5 is forced at wavenumber $k_c = 32$ with Rayleigh friction coefficient $\gamma = 10^{-4}$, corresponding to Re = $5 \times 10^5$. Rayleigh friction suppresses large-scale condensate formation and, combined with the forcing, yields turbulent flows that are far outside the distribution used for training. As such, F4 will be indicative for testing the generalization of the developed model. Furthermore, Rayleigh friction essentially eliminates intermittency, which can make accurate computation of temporal averages very expensive. The numerical solutions are computed on a $8192^2$ grid for DNS to ensure that the dissipation scale is at least 2$\times$ larger than the grid scale. We compute LES on different grids ranging from $32^2$--$1024^2$ for F4 and one $512^2$ grid for the F5 LES models. This resolution corresponds to 
a nondimensional filter scale $\delta \approx 0.08$--$2.6$ and $\delta \approx 0.23$, for flows F4 and F5 respectively. Because the flows are in a statistically stationary state, we compute time averages of relevant observable quantities from one realization of DNS and the various LES models over $\mathcal{O}(10^4)$ eddy turnover times.

\begin{figure}
    \centering
    \begin{subfigure}{0.45\textwidth}
        \includegraphics[width=\textwidth]{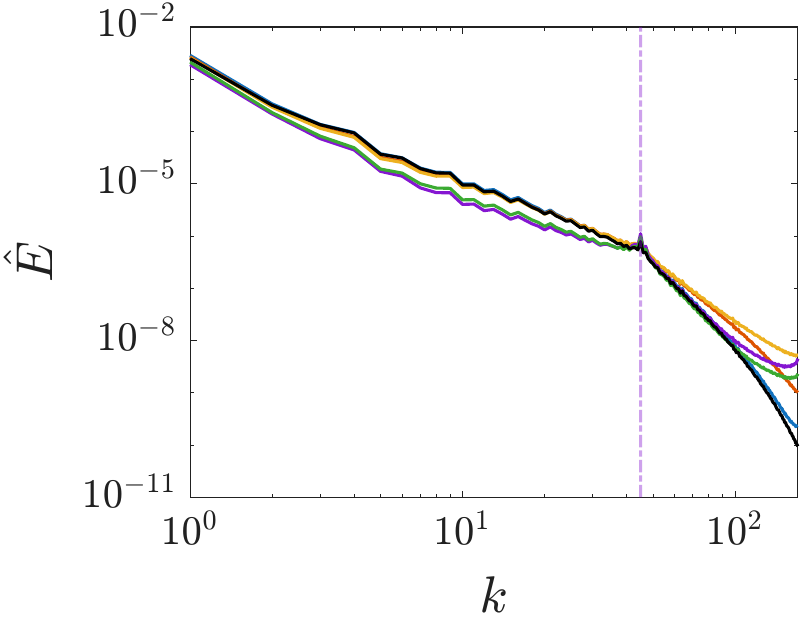}
        \caption{}
    \end{subfigure}
    \hspace{5mm}
    \begin{subfigure}{0.45\textwidth}
        \includegraphics[width=\textwidth]{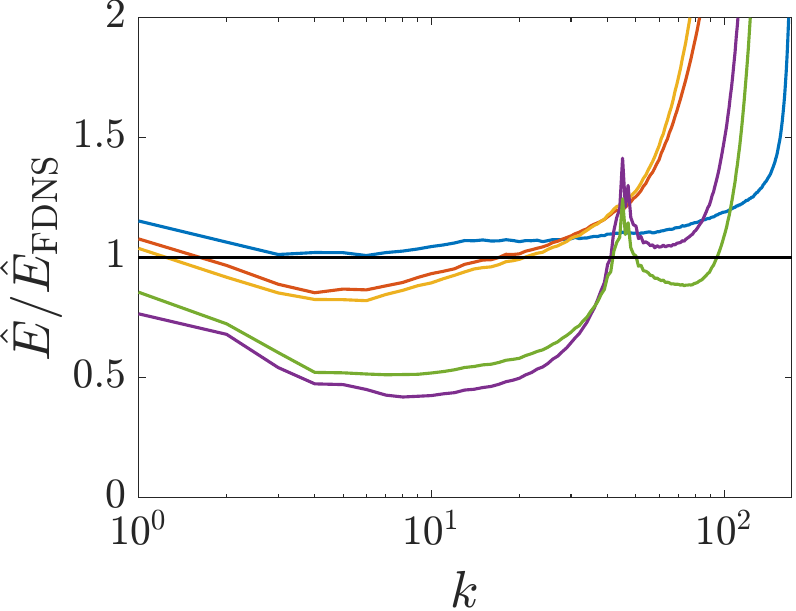}
        \caption{}
    \end{subfigure}
    \caption{Energy spectrum of the flow F5 (a). FDNS (black) is compared with NGMR (blue), NGM4 (orange), similarity (yellow), clipped dynamic mixed (purple), and clipped dynamic Smagorinsky (green). {\color{black}The vertical line shows the forcing frequency.} Respective spectra for each SGS model are normalized by the spectrum of FDNS (b).}
    \label{fig: spectrum}
\end{figure}

\begin{figure}
    \centering
        \begin{subfigure}{0.45\textwidth}
        \centering
        \includegraphics[width=\textwidth]{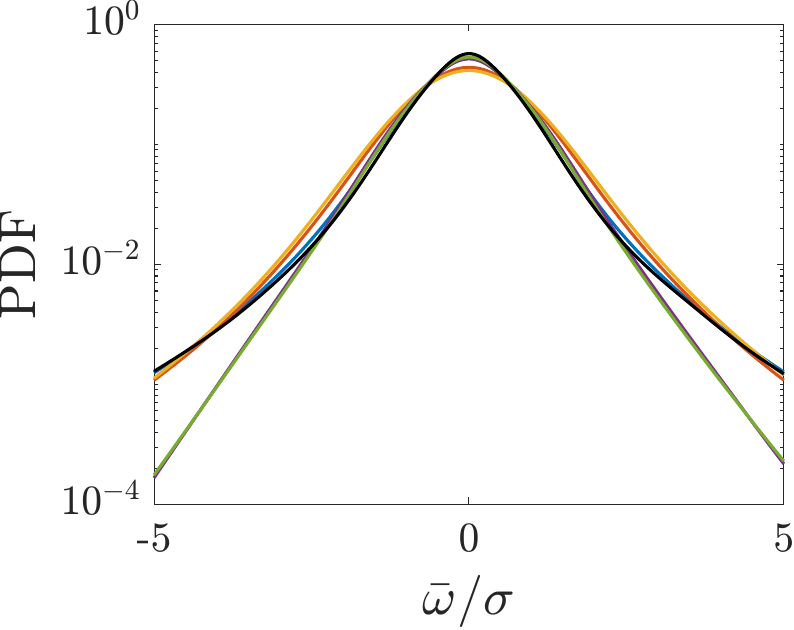}
        \caption{}
    \end{subfigure}
    \hspace{5mm}
    \begin{subfigure}{0.45\textwidth}
        \centering
        \includegraphics[width=\textwidth]{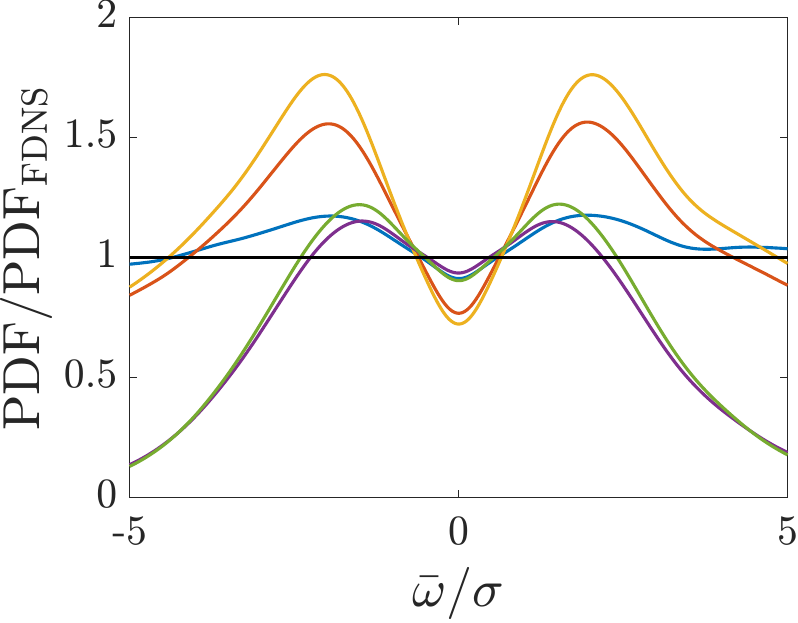}
        \caption{}
    \end{subfigure}
    \caption{Vorticity PDF for the flow F5 (a). FDNS (black) is compared with NGMR (blue), NGM4 (orange), similarity (yellow), clipped dynamic mixed (purple),
and clipped dynamic Smagorinsky (green). PDFs for each SGS model are divided by the PDF of FDNS (b). }
    \label{fig: vorticity PDF}
\end{figure}

To compare the accuracy of different SGS models, we start by computing the time-averaged energy power spectra as well as the probability distribution functions (PDFs) of the pointwise vorticity $\omega$ and local SGS energy flux $\Pi$ for the forced flow F5, which represents the most rigorous and challenging test for any SGS model. 
This high-$\mathrm{Re}$ flow features many small vortices with characteristic size determined by the forcing. It also exhibits both the inverse cascade for $k<k_f$ and the direct cascade for $k>k_f$, where $k_f\approx 45$. As Figure \ref{fig: spectrum} illustrates, the dynamic Smagorinsky and the dynamic mixed model both struggle to correctly describe both cascades. This is likely due to the tendency of both models to over-dissipate energy, as we found in the case of freely-decaying turbulence. On the other hand, the similarity model and NGM4 provide an accurate description of the inverse cascade, while providing the least accurate description of the direct cascade. NGMR is the only model capable of accurately predicting the energy spectrum for both cascades.
For the vorticity PDFs in figure \ref{fig: vorticity PDF}, the dynamic Smagorinsky and the dynamic mixed model provide a reasonably accurate prediction of the central region of the PDF, but fail to describe the tails of the distribution. Similarity and NGM4 is able to noticeably improve matters, as it more accurately captures the tails of the distribution; however they disagree with FDNS everywhere else, particularly in the central, high probability regions. Of the models, NGMR is the only model that is able to accurately describes the entire distribution. 

\begin{figure}
    \centering
    \begin{subfigure}{0.45\textwidth}
        \includegraphics[width=\textwidth]{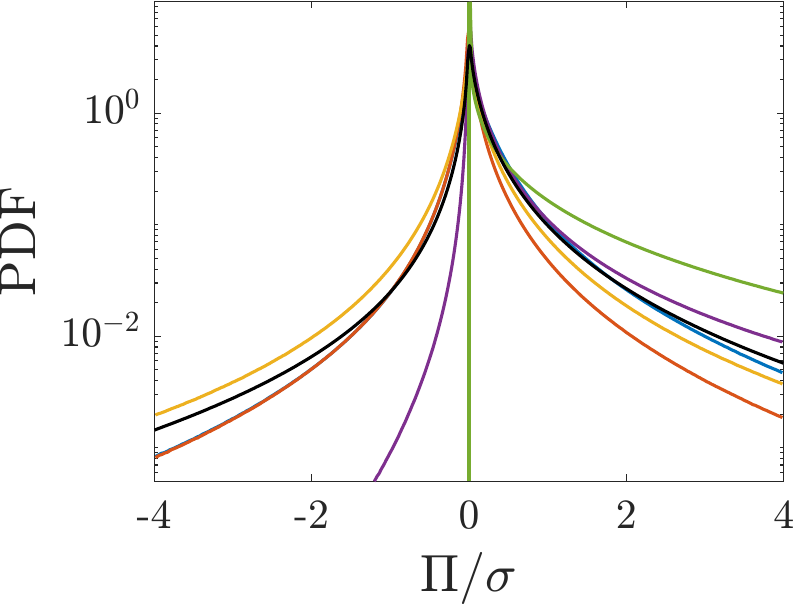}
        \caption{}
    \end{subfigure}
    \hspace{5mm}
    \begin{subfigure}{0.45\textwidth}
        \includegraphics[width=\textwidth]{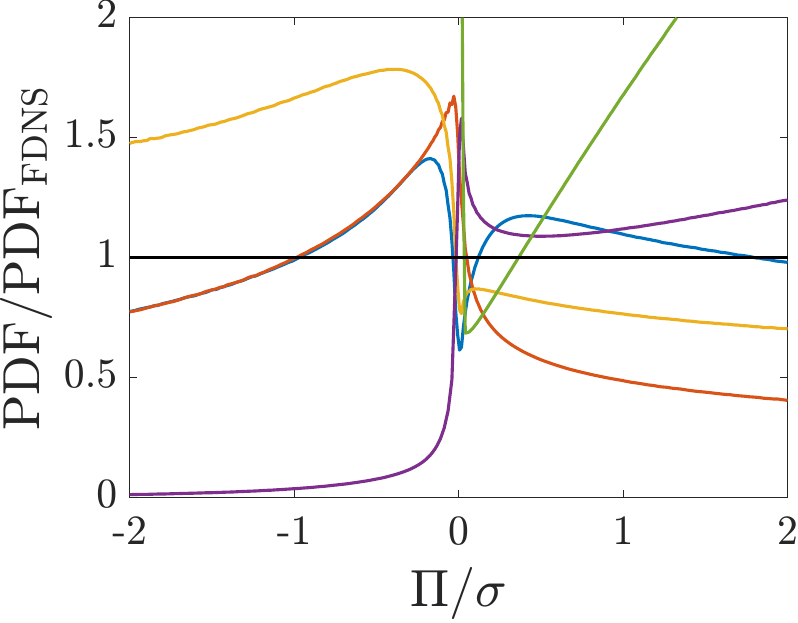}
        \caption{}
    \end{subfigure}
    \caption{Energy flux PDF for the flow F5 (a). FDNS (black) is compared with NGMR (blue), NGM4 (orange), similarity (yellow), clipped dynamic mixed (purple),
 and clipped dynamic Smagorinsky (green). PDFs for each SGS model are divided by the PDF of FDNS at the same filter scale (b).}
    \label{fig: flux PDF}
\end{figure}

The results describing the energy flux PDF in \autoref{fig: flux PDF} are perhaps the most interesting. The distribution is found to be asymmetric for DNS, with dissipation ($\Pi>0$) dominating over backscatter ($\Pi<0$). This is not unexpected, given that in this flow, the energy is primarily dissipated by regular viscosity at the smallest scales rather than Rayleigh friction at all scales. However, backscatter is quite pronounced, which is a requirement for the inverse cascade. Dynamic Smagorinsky and dynamic mixed model both overestimate dissipation and strongly underestimate backscatter. The similarity model exhibits the opposite trend, underestimating dissipation and overestimating backscatter. The corresponding PDF is nearly symmetric. NGM4 yields a reasonably accurate description of backscatter but strongly underestimates dissipation. Finally, NGMR yields the most accurate prediction of both dissipation and backscatter. These findings are qualitatively consistent with our results for the freely decaying flows. 

Having established the superior accuracy of NGMR compared to all the tested models considered here on a variety of benchmarks, we  proceed to look for the limits of its applicability. To allow exploration of a wider range of filter scales, we consider the flow F4 forced at a much lower spatial frequency. 
The {\em a priori} benchmarks shown in \autoref{fig:apri_corr} suggest that NGMR can retain accuracy for $\delta$ as high as 0.6. To quantify the {\em a posteriori} accuracy of this model, we compute two metrics, the energy spectrum and the vorticity PDF, for a range of filter scales. 

\begin{figure}
    \centering
    \begin{subfigure}{0.45\textwidth}
        \centering
        \includegraphics[width=\textwidth]{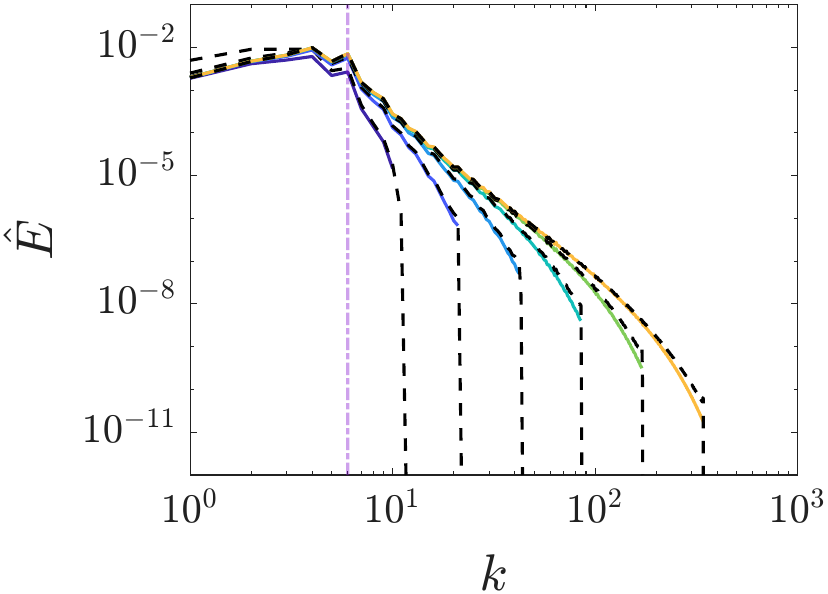}
        \caption{}
    \end{subfigure}
    \hspace{5mm}
    \begin{subfigure}{0.45\textwidth}
        \centering
        \includegraphics[width=\textwidth]{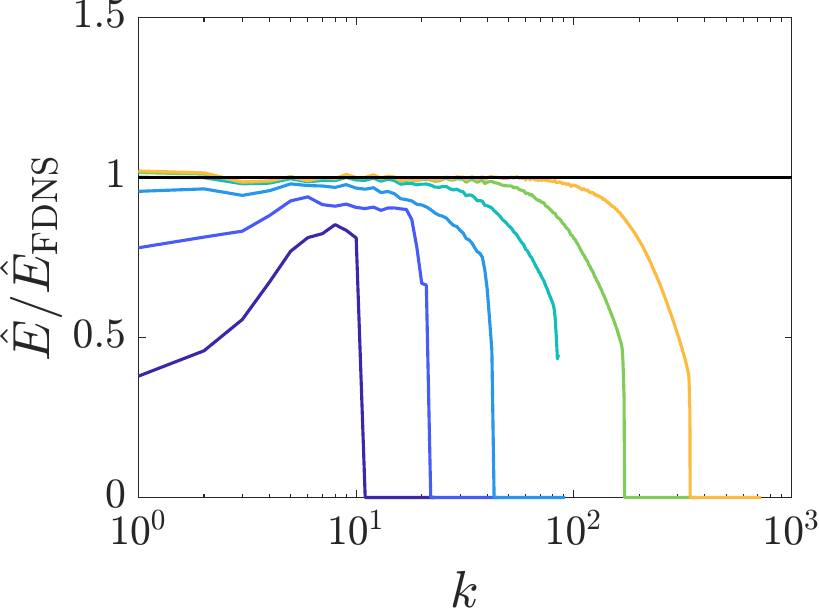}
        \caption{}
    \end{subfigure}

    \caption{The energy spectrum of the flow F4. The color of the lines (violet to yellow) corresponds to nondimensional filter scales $\delta= \{2.401,1.314,0.655,0.328,0.164,0.082\}$.  {\color{black}The vertical line shows the forcing frequency.} Respective spectra for each SGS model are normalized by the spectrum of FDNS at the same filter width (b).}
    \label{fig: Energy low-freq}
\end{figure}

As \autoref{fig: Energy low-freq}(a) shows, NGMR reproduces the energy spectrum of FDNS almost perfectly for the smallest value of the filter scale. Since no SGS model can describe scales smaller than $\Delta$, as the filter scale increases, a gradual deterioration in the accuracy at high frequencies is expected and indeed observed. Filtering, however, also affects lower frequencies. Hence, to make a more objective comparison between the expected energy spectrum $\hat{E}_{\rm{FDNS}}(k)$ and the model prediction $\hat{E}(k)$, we also show their ratio in \autoref{fig: Energy low-freq}(b), with both spectra computed for the same filter scale. The results demonstrate that NGMR remains accurate up to at least $\delta=0.328$.

\begin{figure}
    \centering
     \begin{subfigure}{0.45\textwidth}
        \centering
        \includegraphics[width=\textwidth]{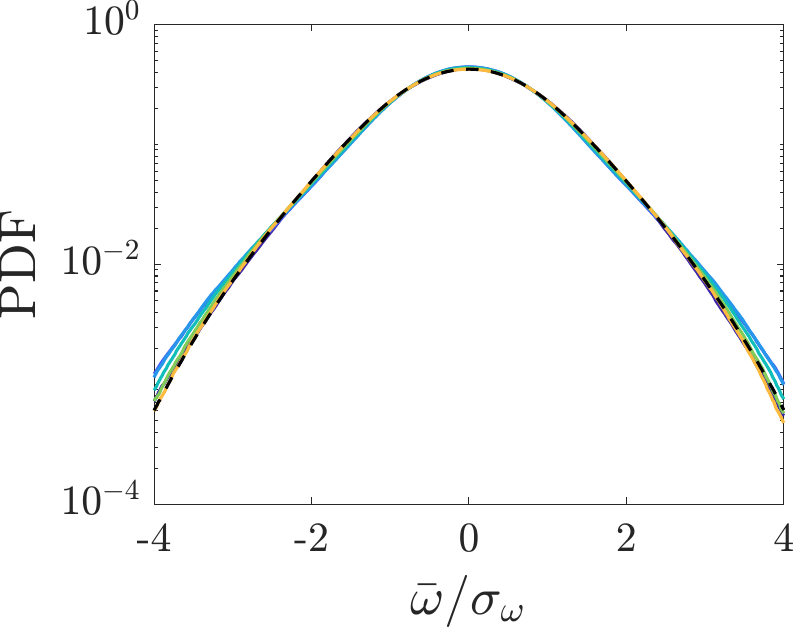}
        \caption{}
    \end{subfigure}
    \hspace{5mm}
    \begin{subfigure}{0.45\textwidth}
        \centering
        \includegraphics[width=\textwidth]{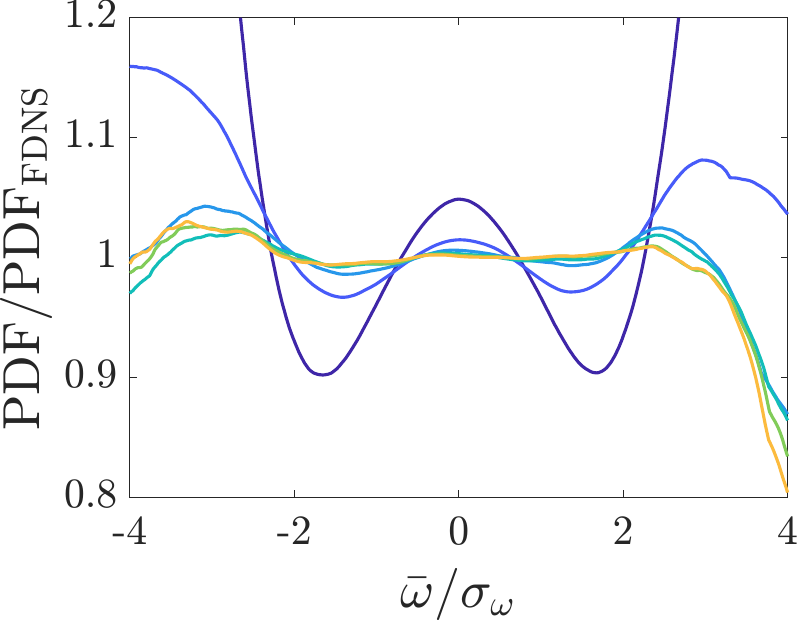}
        \caption{}
    \end{subfigure}
    \caption{The vorticity PDF for the flow F4 (a). The same color scheme as in \autoref{fig: Energy low-freq} is used. Respective PDFs for each SGS model are normalized by the PDF of FDNS at the same filter width (b). }

    \label{fig: Vorticity low-freq}
\end{figure}

A similar conclusion is made by inspecting the vorticity PDF.  \autoref{fig: Vorticity low-freq}(a) shows that NGMR reproduces the PDF obtained using FDNS almost perfectly for the smallest value of the filter scale and quite well even for larger values of $
\delta$. A more objective measure of accuracy is obtained again by inspecting the ratio of the PDFs for the model and FDNS shown in \autoref{fig: Vorticity low-freq}(b). Once again we find that NGMR remains accurate (the ratio stays close to unity) up to at least $\delta=0.655$. Note that the lack of perfect symmetry as well as the deviation of the ratio from unity at the tails of the distribution at small $\delta$, which can be clearly observed in \autoref{fig: Vorticity low-freq}(b), are both due to insufficient statistics rather than model inaccuracy. These deviations are strongly amplified due to the division of one small quantity by another. 

\section{Discussion}\label{sec:discussion}

{\color{black}\subsection{Scale Interaction in 2D Turbulence}}

Perhaps one of the most unexpected results of this study is that, according to \autoref{fig:apri_corr},
an accurate prediction of any one of the three {\it a priori} benchmark quantities (SGS stress tensor $\tau$, the local energy flux $\Pi$, or the net energy flux) does not imply the ability of a SGS model to predict the other two.
For instance, NGM4 consistently yields high-accuracy predictions for the SGS stress tensor but not the local or net energy flux for $\delta\lesssim 0.25$. A similar statement applies in a stronger form to the similarity and dynamic mixed models for $\delta\lesssim0.05$. {\color{black}NGMR, on the other hand, yields highly accurate predictions of the net flux even over the range of filter scales where the predicted local flux deviates noticeably from the true profile.}

\begin{figure}
    \centering
    \begin{subfigure}{0.32\textwidth}
    \includegraphics[width=\textwidth]{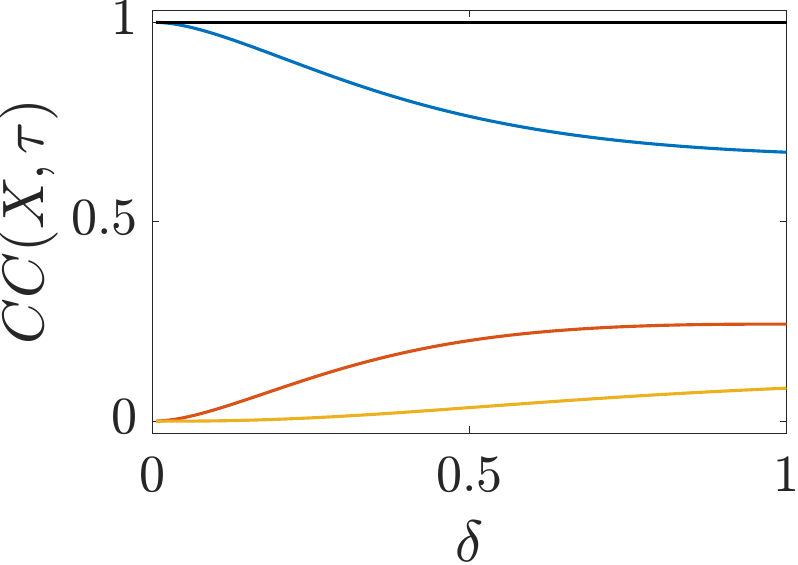}
    \end{subfigure}
    \hspace{1mm}
    \begin{subfigure}{0.32\textwidth}
    \includegraphics[width=\textwidth]{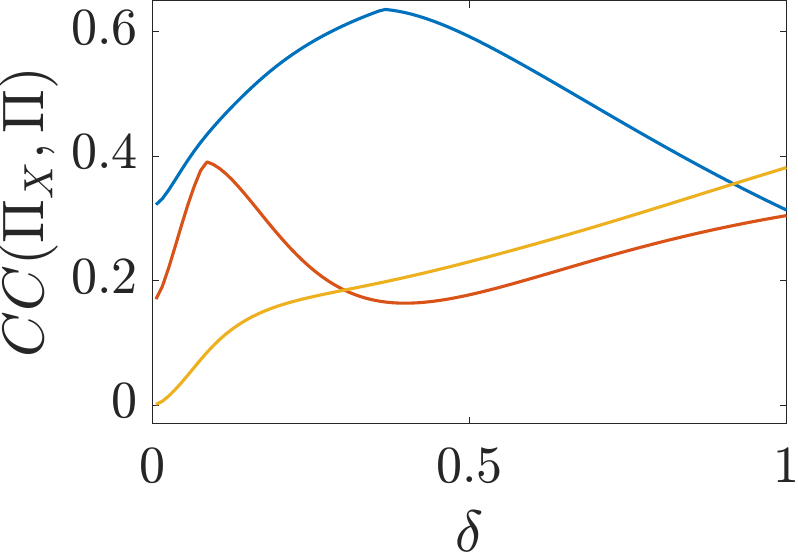}
    \end{subfigure}
    \hspace{1mm}
    \begin{subfigure}{0.32\textwidth}
    \includegraphics[width=\textwidth]{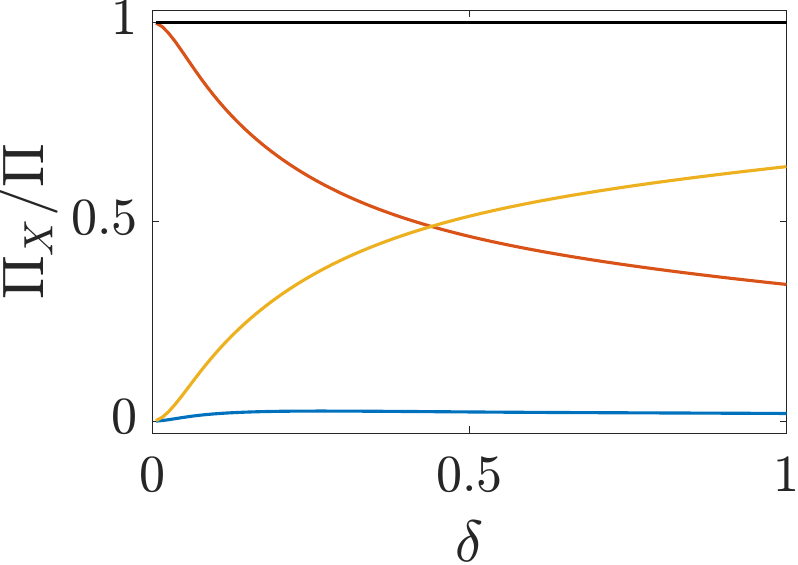}
    \end{subfigure}
    \caption{Comparison of the contributions of the different components $L$, $C$, $R$ (blue, orange, yellow respectively) in the correlation, local flux, and total flux of the SGS stress tensor for the flow F1. 
    The total contribution of $LCR$ is shown in black. }
    \label{fig:lcr_split}
\end{figure}

To understand these results, it is helpful to {\color{black}consider the LCR decomposition of $\tau$ (see \autoref{appendix: Moment Expansion}) and}
quantify the contribution of the tensors $L$, $C$, and $R$ to the SGS stress tensor, the local energy fluxes, and the net energy flux using, respectively, the cross-correlations $\mathrm{CC}(X,\tau)$, $\mathrm{CC}(\Pi_X,\Pi)$ and the ratio $\langle \Pi_X \rangle / \langle \Pi \rangle$, where $\Pi_X \equiv -X_{ij} \bar S_{ij}$ and $X=L,C,R$. Representative results are shown in \autoref{fig:lcr_split} for flow F1. The dynamic Smagorinsky model fails to predict any of the three quantities with any accuracy for any filter scale, so we will instead focus on the other four models. 
As expected, we find that the Leonard stress tensor $L$ yields the dominant contribution to $\tau$ for $\delta\ll1$. As $\delta$ increases, so do the contributions from $C$ and $R$, but $L$ still dominates. This is why the similarity and dynamic mixed models, which focus on modeling $L$ and ignore $C$ and $R$, manage to reproduce $\tau$ somewhat well over a range of (small) filter scales, but not for larger $\delta$. 

We find the opposite situation for the net energy flux. Here, the contribution of $L$ is negligible for all $\delta$, which explains why the similarity model predicts zero flux and the dynamic mixed model fails to predict the flux with any consistency for all filter scales. It is the $C$ tensor that gives the dominant contribution, which is the reason why NGM4, which models both $L$ and $C$ but not $R$, captures the net flux well for $\delta\ll1$. As $\delta$ increases, so does the contribution from $R$, which explains why the predictions of NGM4 quickly deteriorate, while NGMR, which also models $R$, remains highly accurate. Finally, $L$, $C$, and $R$ all yield comparable contributions to the local energy fluxes, which is why NGMR is the only model capable of consistently predicting these fluxes over a wide range of filter scales. These results highlight why it is essential to model all three tensors to predict accurately not only one particular quantity (e.g., net energy dissipation) but a broad collection of physical observables.

{\color{black}\subsection{Modeling Implications}}


Describing backscatter is a difficult challenge that no existing LES model can adequately handle. The {\em a priori} results presented in \autoref{fig:apriori_fluxes} seem to suggest that both NGM4 and NGMR can predict backscatter with comparable accuracy. However, our {\em a posteriori} tests show that the inclusion of the Reynolds stress tensor into the parameterization of the SGS stress tensor is essential for quantitative accuracy. For freely decaying turbulence, NGMR is the only model whose predicted net backscatter tracks that of FDNS for {\color{black}a broad range of Re. The higher-Re} case shown in \autoref{fig:aposteriori_1e-7}(d) is particularly telling: NGMR correctly predicts that backscatter essentially disappears, as it should, when most of the vortices merge to leave a single pair, while all the other models incorrectly predict the net backscatter to asymptote to a nonzero value. In the forced case, the predicted energy flux PDFs for NGMR and NGM4 shown in \autoref{fig: flux PDF} appear indistinguishable for negative values of $\Pi$. A closer inspection shows a noticeable difference at low values of $|\Pi|$. Since the flux PDF is sharply peaked at $\Pi=0$, this difference is strongly amplified, yielding a substantial improvement in the accuracy for NGMR versus NGM4 as illustrated by, e.g., the energy spectra shown in \autoref{fig: spectrum}.  

While NGM4 has the form of a typical LES model, the structure of NGMR is distinctly different, with the Reynolds stress tensor playing the role of a subgrid-scale variable described by its own evolution equation. In this sense, NGMR is more similar to RANS than LES models. Unlike NGMR, most popular RANS models such as the $k$-$\varepsilon$, $k$-$\omega$ and Spalart-Allmaras model \citep{wilcox2008,chien1982,spalart_one-equation_1992} 
use one or more scalar fields describing the subgrid scales and rely on the Boussinesq assumption for the SGS stress tensor. Scalar fields are fundamentally incapable of correctly describing inter-scale fluxes, as fluxes crucially depend on the orientation of the small-scale structures (e.g., vorticity filaments in 2D and vortex sheets in 3D) relative to the expanding and contracting directions of the large-scale flow. NGMR is closer in form to full-Reynolds-stress models such as Wilcox Stress-omega \citep{wilcox2006} and SSG/LRR \citep{eisfeld2016}.
These RANS models describe subgrid scales using the Reynolds stress tensor as well as an additional scalar field ($\omega$) related to turbulent kinetic energy, each described by its own evolution equation similar to our equations \eqref{eq:traceless_r_evo} and \eqref{eq:trace_r_evo}. A notable difference is the opposite sign of the production term in the evolution equation for $R$, which reflects the two-dimensional nature of the flows considered in this study.

{\color{black}\subsection{Model Validation}}

Our {\em a posteriori} results differ considerably from those reported in other studies. For instance, we find a rather significant disagreement between FDNS and NGM4 for both the vorticity PDF and the energy spectra, whereas \citet{jakhar2025} find near-perfect agreement for both metrics. In large part, this is due to the considerably lower values of both $\mathrm{Re}$ and $\delta$ used in their study, which brings us to the issue of proper model validation procedures. Whatever its stated objective, every LES and RANS model is formulated in terms of a parameterization for the SGS stress tensor. Indeed, in order for a model to accurately predict a variety of physical observables rather than, say, just the energy dissipation rate, an accurate parameterization of $\tau$ is essential. The formal moment expansion of $\tau$ is expected to converge, yielding such a parameterization, only for $\delta<O(1)$. On the other hand, as our {\em a priori} benchmarks illustrate, many LES models that yield extremely accurate predictions for $\delta\ll1$ quickly fall apart as $\delta$ increases. Practical considerations dictate that $\delta$ be chosen as large as possible, so both {\em a priori} and {\em a posteriori} benchmarks should be evaluated for $\delta=O(1)$ to obtain a fair evaluation of a model's accuracy.

The choice of $\mathrm{Re}$ is equally important
and depends on the choice of $\delta$. There are two separate sources of dissipation in the momentum equation: molecular viscosity characterized by the energy flux $\Pi_\nu \equiv 2 \nu \bar S_{ij} \bar S_{ij}$ and ``eddy viscosity'' represented by the SGS energy flux $\Pi=-\bar{S}_{ij}\tau_{ij}$. Their ratio defines the Reynolds number $\mathrm{Re}_\Delta$ at the filter scale. Testing a SGS model in the limit of $\mathrm{Re}_\Delta\ll1$, {\color{black}which is common in the LES literature,} can yield highly misleading results. In particular, the stability of a SGS model is highly dependent on the strength of the total dissipation. We found unregularized versions of both NGM4 and the learned correction convolutional neural network model of \citet{kochkov2021} to become numerically unstable at higher $\mathrm{Re}$ (corresponding to $\mathrm{Re}_\Delta\gg1$), while the original studies showed no sign of instability at lower $\mathrm{Re}$ ($\mathrm{Re}_\Delta\ll1$). Regularization appears to be a necessary ingredient for a wide range of structural and hybrid models. The regularization scheme we adopted was found to perform well for pseudo-spectral discretization. Other discretization schemes might require a different regularization to properly balance efficiency, accuracy and numerical stability.

\begin{figure}
    \centering
    \includegraphics[width=0.5\linewidth]{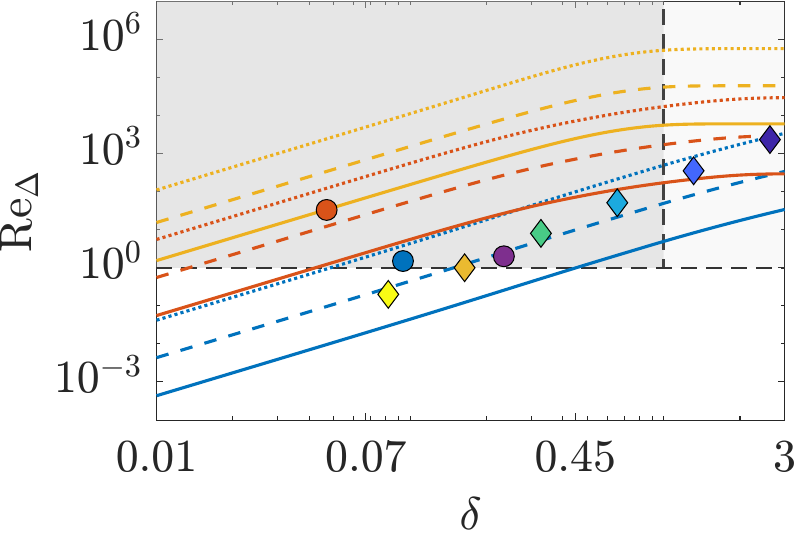}
    \caption{Parameter range for model inference and validation.  The gray box represents the parameter region where model inference and validation should be performed. The training data sets F1 (blue), F2 (orange), and F3 (yellow) are shown as curves. The line style corresponds to the value of viscosity $\nu = 10^{-4}$ (solid), $\nu =10^{-5}$ (dashed) and $\nu=10^{-6}$ (dot-dashed). The symbols represent the data used for {\em a posteriori} tests. The circles correspond to the freely decaying flow F2 with $\nu=10^{-5}$ (blue) and  $\nu=10^{-7}$ (orange) and the forced flow F5 (purple). Diamonds correspond to the forced flow F4 with different filter scales; the same color scheme as in \autoref{fig: Energy low-freq} is used. 
    }
    \label{fig: SGS visc dissipation}
\end{figure}

To properly assess numerical stability, {\em a posteriori} tests should be conducted for $\mathrm{Re}_\Delta>1$, or preferably even $\mathrm{Re}_\Delta\gg1$,  which is, indeed, the intended regime for practical applications of any SGS model. These constraints on $\mathrm{Re}_\Delta$ and $\delta$ apply in equal measure to data-driven model inference, as they ensure that the learned models are sufficiently general, robust, and practically useful. Our training and testing data were generated with constraints on both parameters in mind. \autoref{fig: SGS visc dissipation} provides a summary of where in the $(Re,\delta)$ parameter space our data lie. Both the training data (shown as curves) and testing data (shown as symbols) sample the desired parameter range, $\mathrm{Re}_\Delta>1$ combined with $0<\delta<1$, shown as the gray box, quite well. Together with our results, this convincingly illustrates that NGMR retains both accuracy and stability not just over a wide variety of flow regimes but also over a wide variety of parameter regimes, as a general SGS model should.

{\color{black}\subsection{Limitations and Future Directions}}

We found that a direct parameterization of the turbulent kinetic energy equation in terms of the resolved velocity field using the moment expansion does not significantly hinder the accuracy of our model. However, it is almost certain that the accuracy could be increased further by replacing this parameterization with an evolution equation for $k$ inferred from data which, unlike equation  \eqref{eq:trace_r_evo}, would include terms nonlinear in, or independent of, both $\mathcal{R}$ and $k$, which is a requirement for breaking the scale-invariance of the subgrid-scale variables. 
It is also worth noting that many of the terms in the inferred evolution equation \eqref{eq:R} (and hence equation \eqref{eq:traceless_r_evo}) can be obtained analytically through the moment expansion of $\partial_tR_{ij}$, as shown in \autoref{Appendix: R evo derivation}. In particular, the coefficient $c_1=1$ of the advection term is recovered to all orders in $\Delta$, reflecting the Galilean symmetry of the problem. We also find, to leading order in $\Delta$, the value of two other coefficients: $c_2 = 1$ and $c_3=\nu$. The formal moment expansion yields
\begin{align} 
    \partial_t R_{ij} &= - \bar u_l \nabla_l R_{ij} +
    R_{il} \nabla_l \bar u_{j} + R_{jl} \nabla_l \bar u_{i} + \nu \nabla^2 R_{ij} - \nu\frac{\Delta^6}{3456} \nabla_k\nabla_l \nabla^2 \bar u_i \nabla_k \nabla_l \nabla^2 \bar u_j  \nonumber\\
    &+ 2\nabla_k \nabla^2 \bar u_{(i} M_{kj)}
    {\color{black}\ - 2\gamma R_{ij} 
    + \frac{\Delta^6}{3456}\nabla_k\nabla^2 \bar u_{(i}  \nabla_k \nabla^2\bar f_{j)}} 
    + \mathcal O (\Delta^8),
\end{align}
where $M_{ij}$ is a tensor that cannot be expressed in terms of $\bar{p}$, $\bar{u}_i$, and $R_{ij}$ without introducing higher-order derivatives, which were found to be numerically noisy and led to instabilities in practice. {\color{black}Consequently, the term $2\nabla_k \nabla^2 \bar u_{(i} M_{kj)}$ must also be modeled. 
The two terms representing the contributions of the forcing and Rayleigh friction where not picked up by SPIDER since the training data includes neither physical effect. It is straightforward to include these terms in the NGMR model, which should improve the already exceptional accuracy of the model even further.}
The two viscous terms are negligible at high $\mathrm{Re}$ and can be ignored. The second viscous term was not picked up by SPIDER not only because our training data did not include lower-$\mathrm{Re}$ flows but also because this term is of higher symbolic complexity than what we have included in the search space.

{\color{black}
In this study, we model the tensor field $2\nabla_k \nabla^2 \bar u_{(i} M_{kj)}$ in the form of a linear relaxation term $-\alpha IR_{ij}$; it is mainly this approximation that limits the accuracy of the evolution equation for the Reynolds stress tensor and consequently the entire NGMR model at high $\mathrm{Re}$. Indeed, the spread in the values of the coefficient $\alpha$ produced by regression is a characteristic signature that the accuracy of this parameterization is not very high and could be improved. While it is enticing to look for a better approximation of the prefactor $\alpha I$ in terms of invariants of tensors $\nabla_ju_i$, $\nabla_j\nabla_ku_i$, etc., this does not solve the issue of linearity of the evolution equation which leads to the spurious scaling symmetry. Future work should explore more promising alternatives that abandon the linear approximation altogether. It may be possible to find a parameterization that is not linear in $R$ through sparse regression using higher-complexity libraries. Alternatively, it may be possible to find a more accurate parameterization by introducing additional SGS variables.
}

{\color{black}
Note that none of the models compared in this study are capable of describing the effect of forcing at scales smaller than the filter scale, which is a serious limitation in some applications. Consequently, for the forced flows considered here, the LES resolution is always chosen such that energy-injection scale is resolved. It should be emphasized that the approach presented here can, in principle, be extended to the case where forcing is at scales that are not resolved. Such forcing would appear as a source term in an evolution equation for the Reynolds stress tensor, as shown in \autoref{Appendix: R evo derivation}. This source term also must be modeled and could be deterministic, stochastic, or a combination of both types, depending on the specifics of the forcing. Extending the framework to correctly represent the effect of subgrid-scale forcing is an important direction for future work, especially in systems such as MHD turbulence or geophysical flows, which frequently feature small-scale instabilities or forcing.
}


Although the present study is restricted to two-dimensional incompressible turbulent flows, such flows represent an important and nontrivial testbed for subgrid-scale modeling due to the presence of strong backscatter, exposing modeling deficiencies that may have otherwise remained hidden in fully three-dimensional settings dominated by forward cascades. {\color{black}In particular, we find that to perform well across a broad range of {\it a priori} and {\it a posteriori} benchmarks, a SGS model must} accurately capture the delicate balance between forward and inverse fluxes of various quantities. Moreover, two-dimensional turbulence is not merely a carefully curated test environment, as two-dimensional flows arise naturally in three-dimensional geometries under a variety of conditions, such as rotating, stratified, and vertically confined systems, which are realized in a variety of geophysical and astrophysical settings. {\color{black}In particular, it is straightforward to extend the modeling approach and the model presented here to two-dimensional models of multi-layer quasi-geostrophic flows. Furthermore, the framework presented here should provide a systematic approach to modeling SGS physics in other settings, from three-dimensional fluid turbulence to climate and weather modeling. 
} 

{\color{black}A final point we would like to comment on concerns the treatment of boundaries in more general geometries. The present study considers periodic domains, so boundary effects do not arise. However, applications to wall-bounded flows would require specifying how filtering, higher-order derivative terms, and any auxiliary SGS variables are treated near boundaries. We do not view such complications as a fundamental limitation of the framework. Indeed, SPIDER has been used to infer the boundary conditions from DNS data in a variety of settings such as turbulent channel flow \citep{gurevich_learning_2024} and turbulent Rayleigh-Benard convection \citep{wareing_data-driven_2025}. Extending the present framework to more general geometries, including three-dimensional turbulence and wall-bounded flows, represents a natural and promising extension for future work.}

\section{Conclusion}\label{sec:conclusions}

This study shows that, using a combination of formal analysis and physically informed machine learning, it is possible to systematically construct a numerically stable, symbolic, and hence fully interpretable, subgrid-scale model of fluid turbulence. This model matches or outperforms state-of-the-art LES models on every single metric, {\em a priori} or {\em a posteriori}, for a wide range of two-dimensional, incompressible turbulent flows. Moreover, this model retains accuracy over a much wider range of filter scales and Reynolds numbers. Equally important, the mathematical structure of the model highlights and explains the deficiencies of existing LES models, charting a path towards building even more accurate reduced descriptions of turbulence and other multiscale phenomena in both two and three spatial dimensions. Extension of this model to three-dimensional and/or compressible fluid turbulence is a natural next step.

\medskip\noindent

\medskip\noindent
{\bf Acknowledgements.} The authors thank Matthew Golden for helpful discussions and insights related to the SPIDER framework and model discovery. This research was supported in part through research resources and services provided by the Partnership for an Advanced Computing Environment (PACE) at the Georgia Institute of Technology, Atlanta, Georgia, USA. The authors used generative artificial intelligence tools for minor programming assistance (e.g. code debugging). All scientific content and conclusions were developed and verified by the authors, who take full responsibility for the manuscript.

\medskip\noindent
{\bf Funding.} {\color{black}This material is based upon work supported by the Defense Advanced
Research Projects Agency (DARPA) under Agreement No. HR00112590005. The views and conclusions
contained in this document are those of the authors and should not be interpreted as representing the
official policies, either expressed or implied, of the U.S. Government.}

\medskip\noindent
{\bf Declaration of Interests.} The authors report no conflict of interest.

\medskip\noindent
{\bf Data availability statement.} The data that support the findings of this study are openly available in GitHub at https://github.com/cdggt/jax-cfd-sgs/



\clearpage
\appendix

\section{Computational details \& DNS}\label{Appendix}
\subsection{Numerical Solver}\label{Appendix: Numerics}
All DNS simulations were performed using the JAX-CFD framework \citep{jax-cfd}, using a pseudospectral solver for the incompressible 2D Navier--Stokes equation in vorticity--streamfunction formulation on a doubly periodic square domain ($2\pi \times 2\pi$). The full numerical pipeline, including data generation, filtering, and analysis scripts, is available in an open-source repository \citep{jax-cfd-sgs}. The grid resolution was selected such that it is at least twice as fine (accounting for dealiasing) as the dissipation scale, which corresponds to $\mathrm{Re}_\Delta=1$. To ensure that numerical solutions are fully resolved, we applied SPIDER to each DNS simulation and verified that the governing equations (continuity and Navier--Stokes) hold to within machine precision {\color{black}in weak form. This is essential for data-driven inference of the SGS model based on this framework}. 

The time integration employs a Crank--Nicolson scheme for the viscous terms and a fourth-order Runge--Kutta (RK4) scheme for the nonlinear terms, with the timestep satisfying $\mathrm{CFL}=u_{\max}\Delta t/\Delta x \le 0.3$ in all simulations. The CFL condition is computed based on the maximum resolved velocity magnitude. The timestep is fixed within each run. All viscous terms are treated implicitly. All computations were performed in double precision to eliminate any potential accumulation of numerical error; single-precision tests yielded quantitatively indistinguishable results. Nonlinear terms were evaluated pseudo-spectrally by computing products in physical space followed by Fourier transformation, with aliasing errors removed using the standard $2/3$ truncation rule. 

For the {\em a posteriori} evaluations, each closure scheme was embedded directly into the JAX-CFD solver, which was validated using an equivalent MATLAB implementation that employs an identical numerical scheme. All LES models are solved in the velocity formulation, with a Leray projection used to enforce incompressibility via a spectral pressure projection step. Subgrid-scale terms are treated implicitly wherever numerically feasible and beneficial for stability; remaining terms are treated explicitly.

{\color{black}
Note that our numerical implementation does not attempt to retain symmetry equivariance explicitly, which is a property of the continuous SGS model. A finite-dimensional numerical implementation need not preserve the full continuous symmetry group exactly: filtering, projection, dealiasing, and the discrete representation of derivatives can weakly break these symmetries in a scheme- and resolution-dependent way. Thus, our pseudospectral solver should be viewed as a discretization of a continuum-equivariant model, not as an explicitly equivariant numerical scheme.
}

\subsection{Initial Conditions}
\begin{figure}
    \centering
    \begin{subfigure}{0.33\textwidth}
        \centering
        \includegraphics[width=\textwidth]{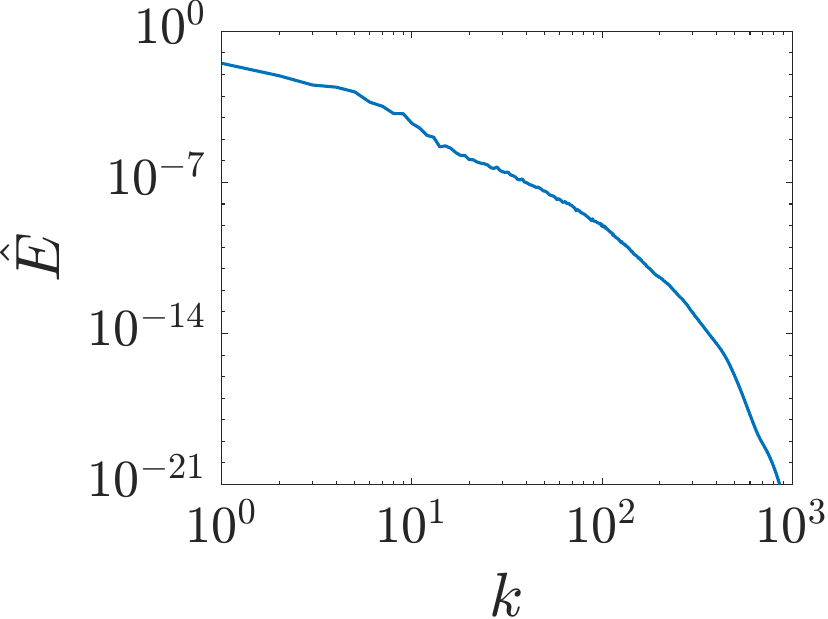}
        \caption{}
    \end{subfigure}
    \begin{subfigure}{0.32\textwidth}
        \centering
        \includegraphics[width=\textwidth]{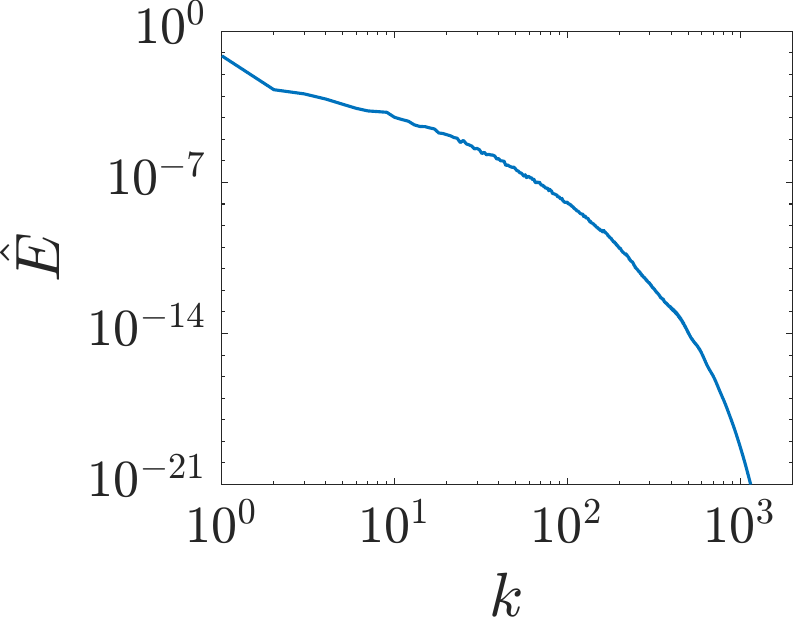}
        \caption{}
    \end{subfigure}
    \begin{subfigure}{0.33\textwidth}
        \centering
        \includegraphics[width=\textwidth]{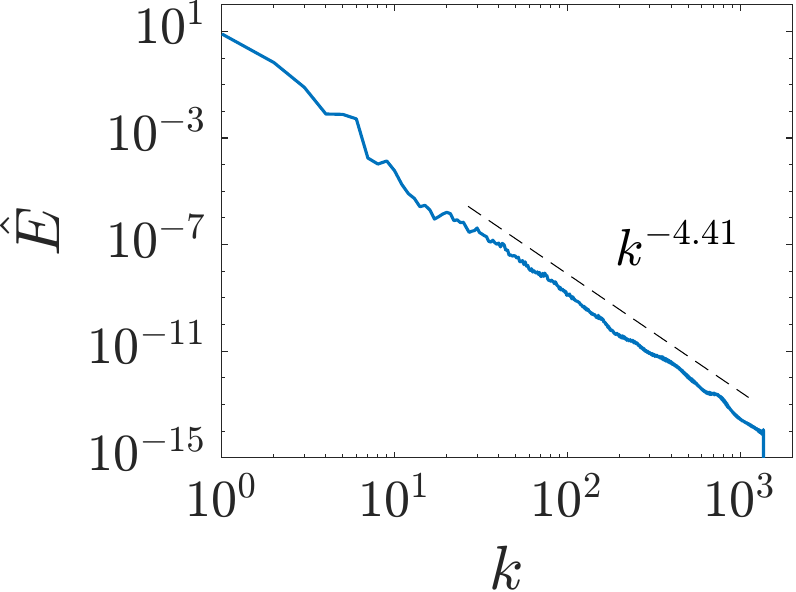}
        \caption{}
    \end{subfigure}

    \caption{
    Energy spectrum of the initial condition for DNS at $\nu = 10^{-5}$ of flows F1 (a), F2 (b) and F3 (c). 
    }
    \label{fig: Freedecay IC Spectrum}
\end{figure}

The {\em a priori} analysis was performed with DNS data representative of different canonical turbulent regimes generated at $\nu = 10^{-4}$, $10^{-5}$, and $10^{-6}$ using grids of $2048^2$, $4096^2$, and $4096^2$, respectively. The resulting data sets reproduce the expected inertial-range scaling characteristic of two-dimensional turbulence. Zero spatial mean vorticity was enforced in every case. Flows F1 and F3 were initialized from random vorticity fields and evolved under a steady checkerboard forcing \eqref{eq:checkerboard}. Each flow was integrated until the total energy reached a statistically stationary state. Flow F1 was forced at an intermediate wavenumber $k_c=32$, allowing an inverse energy cascade to develop, while still leaving sufficient spectral separation to permit filtering over a wide range of filter scales $\Delta$ below the forcing scale. Flow F3 was forced at a low wavenumber $k_c=4$, allowing a direct cascade to form. Flow F2 was obtained by further evolving flow F1 after turning off the forcing, allowing vortex mergers to proceed and yielding a flow dominated by a small number of large coherent vortices. No linear (Rayleigh) friction, hyperviscosity, or hypoviscosity was used in generating the three initial conditions or any of the nine DNS data sets employed for model inference.
The corresponding energy spectra are shown in \autoref{fig: Freedecay IC Spectrum} for the intermediate value of viscosity. Flow F3 features a wide inertial range with a scaling exponent of $\alpha=-4.41$, consistent with expectations \citep{zhigunov2023,reynoso2024}. Flow F2 is transient and displays no inertial range, also as expected. Finally, flow F1 too lacks a clear inertial range due to strong intermittency characteristic of the lack of Rayleigh friction.   

\begin{figure}
    \centering
    \begin{subfigure}{0.45\textwidth}
        \centering
        \includegraphics[width=\textwidth]{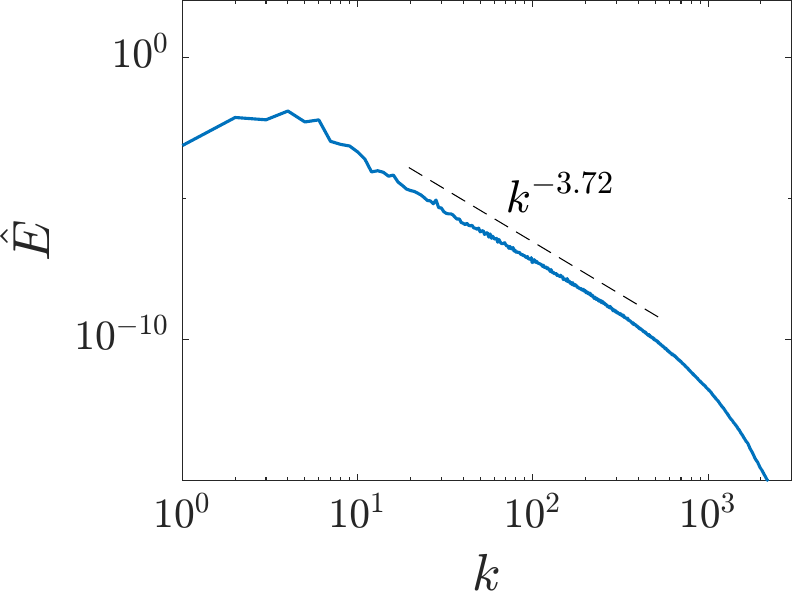}
        \caption{}
    \end{subfigure}
    \hspace{5mm}
    \begin{subfigure}{0.45\textwidth}
        \centering
        \includegraphics[width=\textwidth]{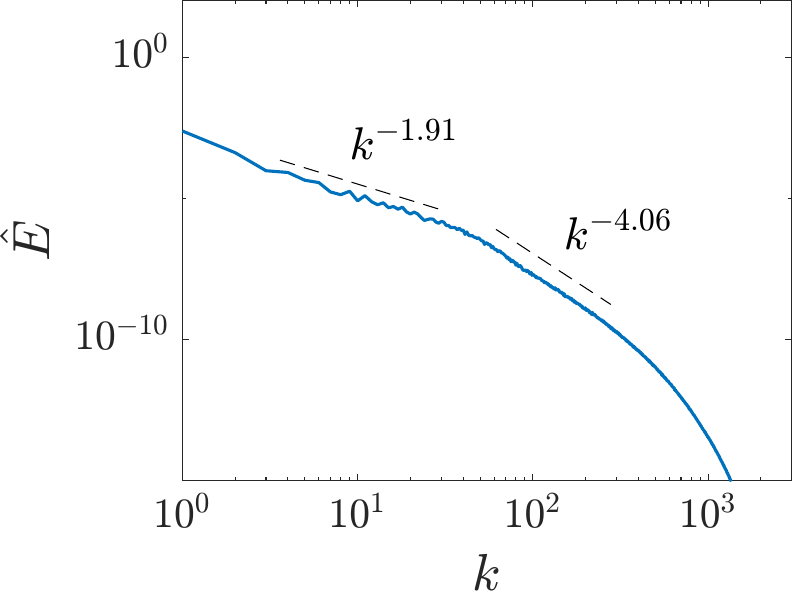}
        \caption{}
    \end{subfigure}

    \caption{Energy spectrum of the initial condition for DNS of flows F4 (a) and F5 (b). 
    }
    \label{fig: Forced IC Spectrum}
\end{figure}

Flow F4 is forced at wavenumber $k_c = 4$ with Rayleigh friction with coefficient $\gamma = 0.05$, while flow F5 is forced at wavenumber $k_c = 32$ and includes Rayleigh friction with $\gamma = 10^{-4}$. The initial conditions were generated by evolving both flows until the total energy reached a statistically stationary state. In contrast to the freely decaying flows, both the forcing and friction were maintained while generating the {\em a posteriori} benchmarking data. The energy spectra of the corresponding initial conditions are shown in \autoref{fig: Forced IC Spectrum}. Rayleigh friction essentially eliminates intermittency, yielding clean power law scaling for each cascade. Flow F4 exhibits a direct cascade only, with a wide inertial range and scaling exponent of \textcolor{black}{ $\beta=-3.72$}. Flow F5 exhibits both an inverse cascade (with scaling exponent \textcolor{black}{$\beta=-1.91$}) and a direct cascade (with scaling exponent \textcolor{black}{$\beta=-4.06$}). These results are again consistent with expectations \citep{zhu2024}. Note that the scaling exponents predicted by the classical Kraichnan--Leigh--Batchelor theory (\textcolor{black}{$\beta=-5/3$} for the inverse cascade and \textcolor{black}{$\beta=-3$} for the direct cascade{\color{black}, derived under the assumption of a constant energy or enstrophy flux, respectively}) describe only highly idealized flows with hypo- and hyperviscosity, no Rayleigh friction and, most importantly, no coherent structures present.  

LES simulations were initialized using the corresponding FDNS solutions and advanced in time using the pseudospectral integration procedure described in Appendix \ref{Appendix: Numerics}. All filtering was performed in Fourier space using a Gaussian filter with second moment $\sigma^2 = \Delta^2/12$, followed by Fourier projection onto the coarse LES grid and combined with 2/3 dealiasing. The filter scale was set at $\Delta = 6\pi / N_{\mathrm{LES}}$, where $N_{\mathrm{LES}}$ is the number of grid points in each spatial direction. 

\section{LES models}\label{Appendix: Models}
To evaluate the performance of NGMR, we benchmark it against NGM4 and three representative subgrid-scale (SGS) models most commonly used in large-eddy simulation of incompressible turbulence \citep{meneveau2000}. The selected models span the primary main types of subgrid-scale modeling: functional (Dynamic Smagorinsky), structural (Similarity), hybrid structural–functional (dynamic Mixed), and gradient-based closures (NGM4). Together, these models thereby provide a representative and increasingly stringent set of baselines for assessing the performance of any new SGS model. 

Implementations are identical across the JAX-CFD and MATLAB solvers, and aside from the SGS parameterization itself, all models employ the same numerical discretization, filtering procedures, and time-stepping described in Appendix~\ref{Appendix: Numerics}. Spatial derivatives and filtering operations are performed in Fourier space; consequently, derivatives commute exactly with the {\color{black}Gaussian filter (but not the sharp spectral filter that corresponds to discretization)}. Each model is summarized below in its canonical formulation following the notation and conventions of \citet{meneveau2000}.

\textit{Similarity model ---} The Similarity model \citep{Bardina1980} represents the SGS tensor by the difference between two successive filter levels:
\begin{equation}
    \tau_{ij}^{\mathrm{ Sim}}
    = C_{\mathrm{ Sim}}\;\left(\;\widetilde{\bar{u}_i \bar{u_j}} -\widetilde{\bar{u}}_i\widetilde{\bar{u}}_j
    \right),
\end{equation}
where tilde describes a test filter of width \(\alpha\Delta\). Following the original formulation of \citep{Bardina1980} we set $C_{\mathrm{ Sim}}= \alpha = 1$, making the similarity SGS tensor precisely equal to the standard Leonard stress tensor.  

\textit{Dynamic Smagorinsky model ---} The Dynamic Smagorinsky model \citep{germano_dynamic_1991, lilly_proposed_1992} assumes an eddy-viscosity form for the subgrid stress tensor
\begin{equation}
    \tau_{ij}^{\mathrm{DSmag}} = - 2 \nu_T \bar{S}_{ij}, \quad \nu_T = (C_s\Delta)^2|\bar{S}|.
\end{equation}
The Smagorinsky coefficient $C_s^2$ is computed dynamically from the Germano identity
\begin{equation}
    L_{ij} = \widetilde{\bar{u}_i \bar{u}_j} - \tilde{\bar{u}}_i \tilde{\bar{u}}_j,
\end{equation}
where tilde describes a test filter of width $\alpha\Delta$. We follow the convention of $\alpha = 2$ \citep{germano_dynamic_1991, lilly_proposed_1992}. Applying least-squares minimization yields
\begin{equation}
    C_s^2 = \frac{ L_{ij}M_{ij}}{ M_{ij}M_{ij}}, \quad M_{ij}= -2\Delta^2 \left(|\widetilde{\bar{S}}|\widetilde{\bar{S}}_{ij} -\widetilde{|\bar{S}|\bar{S}_{ij}}\right)
\end{equation}
For our {\em a posteriori} results, we follow common LES practice by clipping the eddy-viscosity $\nu_T$ in order to enforce a positive–definite eddy viscosity, which ensures a purely dissipative baseline model for purposes of stability. 

\textit{Dynamic mixed model ---} The Dynamic mixed model \citep{Bardina1980, zang1993} is a combination of the Dynamic Smagorinsky and Similarity models 
\begin{equation}
    \tau_{ij}^{\mathrm{DM}} =  \tau_{ij}^{\mathrm{Sim}} + \tau_{ij}^{\mathrm{DSmag}} = \overline{\bar{u}_i\bar{u}_j}-\bar{\bar{u}}_i\bar{\bar{u}}_j - 2 \nu_T \bar{S}_{ij},
\end{equation}
where $\nu_T = (C_s\Delta)^2|\bar{S}|$ is obtained from the dynamic procedure described previously. We again employ clipping of $\nu_T$ in all of our {\em a posteriori} simulations. 

\textit{NGM4 ---} This fourth-order nonlinear gradient model \citep{jakhar2025} corresponds to the truncation of the moment expansion of $\tau$ at fourth order in $\Delta$: 
\begin{equation}
    \tau_{ij}^{\mathrm{NGM4}} =  \tfrac{\Delta^2}{12}(\nabla_k \bar u_i) (\nabla_k \bar u_j) + \tfrac{\Delta^4}{288}(\nabla_k\nabla_m\bar{u}_i)(\nabla_k\nabla_m\bar{u}_j)
\end{equation}
The first term corresponds to the second–order nonlinear gradient model, also known as the Clark model \citep{jakhar2024, clark1979}. NGM4 is of particular interest because it represents NGMR without the Reynolds stress tensor $R_{ij}$. 

\section{SGS stress tensor decomposition and moment expansion} \label{appendix: Moment Expansion}

{\color{black}To ensure the coarse-grained description remains accurate,} the loss of information associated with using a coarse grid must be compensated for by introducing new variable(s) that describe subgrid scales.
In order to identify {\color{black}and interpret the proper SGS variables, it is fruitful, although not strictly necessary,} to consider the LCR decomposition of $\tau$ \citep{clark_evaluation_1979,germano_proposal_1986}, into three Galilean-invariant components $L$, $C$, and $R$, given by
\begin{subequations}\label{eq:LCR}
\begin{align}
  \tau_{ij} & = L_{ij} + C_{ij} + R_{ij},
   \\[3pt]
  L_{ij} & =  \overline{\bar{u}_i \bar{u}_j} - \overline{\bar{u}}_i  \overline{\bar{u}}_j, 
   \\[3pt]
  C_{ij} & = \overline{\bar{u}_i u'_j + u'_i \bar{u}_j} -  \overline{\bar{u}}_i \bar{u'_j} -\bar{u'}_i \overline{\bar{u}}_j,
   \\[3pt]
  R_{ij} & = \overline{ u'_i u'_j } - \overline{u'_i}\, \overline{u'_j}. \label{eq: R def}
\end{align}
\end{subequations}
For invertible filters {\color{black}and a suitably small filter scale}, it is possible to represent $L$, $C$, and $R$ as a function of the resolved velocity $\bar{\textbf{u}}$ using the moment expansion. For a spatially homogeneous and isotropic filter with finite moments (as is the one adopted in this paper), the filtering operator \eqref{filteroperator} can be written in terms of the gradient operator \citep{sagaut_large_2006}
\begin{equation} \label{differential_filter}
     \bar{\phi}=
     \left(1  
    + \frac{1}{2}\Sigma_{ij}^{(1)} \nabla_i\nabla_j
    +\frac{1}{24}\Sigma_{ijkl}^{(2)} \nabla_i\nabla_j\nabla_k\nabla_l + \cdots\right)\phi, 
\end{equation}
where $\Sigma^{(n)}=O(\Delta^{2n})$ are tensor coefficients determined by the moments of the filter kernel $G$. Note that the right-hand-side of Equation \eqref{differential_filter} is a power series in the nondimensional parameter $\delta \equiv \Delta/\ell_i$,
where $\ell_i$ is the characteristic length scale of the flow, 
with subsequent terms providing progressively smaller corrections for $\delta\ll1$. 
{\color{black}In the continuum limit,} this series can be inverted to express $\phi$ in terms of $\bar{\phi}$. For a Gaussian kernel with second moment $\Delta^2/12$ as used here, one obtains
\begin{equation}\label{filter_inverse}
    \phi = \bar{\phi}
    -\frac{\Delta^2}{24}\nabla^2 \bar{\phi}
    + \frac{\Delta^4}{1152}\nabla^4\bar{\phi}
    + \mathcal{O}(\Delta^6).
\end{equation}

Applying the moment expansion to the SGS stress tensor, one finds
\begin{eqnarray} \label{LC_ngm_approx}
    \tau_{ij} = \frac{\Delta^2}{12}  (\nabla_k \bar u_i) (\nabla_k \bar u_j) + \frac{\Delta^4}{288}(\nabla_k\nabla_l\bar u_i)(\nabla_k\nabla_l \bar u_j)+\mathcal{O}(\Delta^6),
\end{eqnarray}
{\color{black}which coincides with our data-driven parameterization \eqref{eq:ngm4}.}
One can also apply the moment expansion to the component tensors and find that $L=O(\delta^2)$, $C=O(\delta^4)$, and $R=O(\delta^6)$, so that $L$ provides the leading-order contribution to $\tau$, with $C$ and $R$ representing progressively smaller corrections. {\color{black}Specifically,
\begin{subequations}
\begin{align}
    L_{ij} &= \frac{\Delta^2}{12}  (\nabla_k \bar u_i) (\nabla_k \bar u_j) + \mathcal{O}(\Delta^4),\\
    L_{ij}+C_{ij} &= \frac{\Delta^2}{12}  (\nabla_k \bar u_i) (\nabla_k \bar u_j) + \frac{\Delta^4}{288}(\nabla_k\nabla_l\bar u_i)(\nabla_k\nabla_l \bar u_j)+\mathcal{O}(\Delta^6)
\end{align}
\end{subequations}
and
\begin{align}
    \tau=L+C+O(\delta^6),
\end{align}
so it makes sense to interpret the three terms in the parameterization \eqref{eq:tauR} as analogues of, respectively, $L$, $C$, and $R$.}
Note that carrying out expansions such as \eqref{filter_inverse} and \eqref{LC_ngm_approx} to higher orders quickly becomes meaningless on a coarse grid due to discretization errors, leading to a loss of accuracy and, ultimately, divergence of the series. {\color{black}This is a consequence of the sharp spectral filter being non-invertible.}
Indeed, an LES model including terms up to $O(\delta^6)$, referred to as NGM6, was found to give no improvement in accuracy compared to NGM4 \citep{jakhar2025}. 


{\color{black}
An interesting property of the leading order truncation $\tau=\tau^{(2)}\propto (\nabla\bar{\mathbf u})(\nabla\bar{\mathbf u})^T$ is that the corresponding energy flux $\Pi^{(2)}$ vanishes identically. To see this, we write the resolved velocity gradient as
\begin{equation}\label{eq:SOmega}
    \nabla\bar{\mathbf u}=\bar S+\bar \Omega,
\end{equation}
so that
\begin{equation}
    \Pi^{(2)}
    =
    -\tau_{ij}^{(2)}\bar S_{ij}
    \propto\operatorname{tr}\left[
    (\nabla\bar{\mathbf u})(\nabla\bar{\mathbf u})^T\bar S
    \right].
\end{equation}
Since $(\nabla\bar{\mathbf u})^T=\bar S-\bar \Omega$, we have
\begin{align}
    \operatorname{tr}\left[
    (\nabla\bar{\mathbf u})(\nabla\bar{\mathbf u})^T\bar S
    \right]
    &=
    \operatorname{tr}\left[
    (\bar S+\bar \Omega)(\bar S-\bar \Omega)\bar S
    \right] \\
    &=
    \operatorname{tr}(\bar S^3)
    -
    \operatorname{tr}(\bar \Omega^2\bar S),
\end{align}
where all other terms cancel by the cyclic property of the trace.

For an incompressible flow, $\operatorname{tr}\bar S=\nabla\cdot\bar{\mathbf u}=0$. Since $\bar S$ is then a traceless $2\times2$ symmetric tensor, its eigenvalues are $\lambda$ and $-\lambda$, so $\operatorname{tr}(\bar S^3)=0$. Moreover, any two-dimensional antisymmetric tensor satisfies $\Omega^2\propto \mathbbm{1}$, so
\begin{equation}
    \operatorname{tr}(\bar\Omega^2\bar S)
    \propto
    \operatorname{tr}\bar S
    =
    0
\end{equation}
and therefore
\begin{equation}
    \Pi^{(2)}=0
\end{equation}
for any two-dimensional incompressible velocity field.

Thus, although $\tau^{(2)}$ can be highly correlated with the true $\tau$, it is energetically inactive in 2D and cannot describe the flux of energy between scales regardless of direction. This cancellation is special to two dimensions; in 3D, incompressibility does not force either $\operatorname{tr}(\bar S^3)$ or $\operatorname{tr}(\Omega^2\bar S)$ to vanish.
}

\section{Evolution equation for the Reynolds stress tensor}\label{Appendix: R evo derivation}

This section discusses the analytical derivation of the evolution equation for the Reynolds stress tensor $R$.
In the continuum limit, this tensor can be approximated with the moment expansion 
\begin{equation} \label{R_series_def}
    R_{ij} = \sum \limits_{n=1}^\infty c_n\Delta^{2n} (D_n^a \bar u_i) (D_n^a \bar u_j),
\end{equation}
where $D_n^a = \nabla_k\nabla_l \cdots \nabla_m$ is a tensorial differential operator, $n$ indicates its order, $a$ is a shorthand multi-index, and the numerical constants $c_n$ absorb the corresponding moments of the Gaussian filter. To leading order in $\Delta$,
\begin{eqnarray} \label{leading_order_R}
    R_{ij} = \frac{\Delta^6}{6912}(\nabla_k\nabla^2 \bar u_i) (\nabla_k \nabla^2 \bar u_j) + \mathcal O (\Delta^8).
\end{eqnarray}
Taking the time derivative, we find
\begin{eqnarray}
    \partial_t R_{ij} = \frac{\Delta^6}{3456}(\nabla_k \nabla^2 \bar u_{(i})( \nabla_k \nabla^2 \partial_t \bar u_{j)}),
\end{eqnarray}
where ``Sym.'' denotes term(s) required to make the expression on the right-hand side symmetric in the indices $i$ and $j$. 
Substituting equation \eqref{eq:f_ns} yields
\begin{eqnarray}
    \partial_t R_{ij} = \frac{\Delta^6}{3456} (\nabla_k \nabla^2 \bar u_{(i} )\nabla_k \nabla^2 \left(- \bar u_l \nabla_l \bar u_{j)} - \nabla_{j)} \bar p+\nu \nabla^2 \bar u_{j)} + \nabla_l \tau_{lj)} 
    {\color{black}\ - \gamma \bar{u}_{j)} + \bar{f}_{j)}}\right),
\end{eqnarray}
where the last term involving the subgrid stress tensor is higher-order in $\Delta$ and can be ignored. Collecting terms and using the filtered pressure-Poisson equation to remove the dependence on $\bar{p}$, we obtain
\begin{align}
    \partial_t R_{ij} &= \frac{\Delta^6}{3456} \nabla_k \nabla^2 \bar u_{(i} \big[ \nabla_k \nabla_{j)} (\nabla_l \bar u_m \nabla_m \bar u_l) - \nabla_k \nabla^2 \left(\bar u_l \nabla_l \bar u_{j)}  \right) \\
    &+ \nu  \nabla^4 \bar u_{j)} {\color{black}\ - \gamma \nabla_k \nabla^2\bar{u}_{j)} + \nabla_k \nabla^2\bar{f}_{j)}} \big] + \mathcal{O}(\Delta^8).
\end{align}
Using the tensor identity (which is only valid in 2D)
\begin{equation}
    (\nabla_k \nabla_j (\nabla_i u_l - \nabla_l u_i))(\nabla_l u_j ) = (\nabla_k \nabla^2 u_l) (\nabla_l u_i),
    \label{2D_equality}
\end{equation}
one can rewrite the series in terms of $R_{ij}$ by using equation \eqref{leading_order_R} one more time, yielding
\begin{align} \label{eq:R_NGM}
    \partial_t R_{ij} &= - \bar u_l \nabla_l R_{ij} +
    R_{il} \nabla_l \bar u_{j} + R_{jl} \nabla_l \bar u_{i} + \nu \nabla^2 R_{ij} - \nu\frac{\Delta^6}{3456} \nabla_k\nabla_l \nabla^2 \bar u_i \nabla_k \nabla_l \nabla^2 \bar u_j  \nonumber\\
    &+ 2\nabla_k \nabla^2 \bar u_{(i} M_{kj)}
    {\color{black}\ - 2\gamma R_{ij} 
    + F_{ij}} 
    + \mathcal O (\Delta^8),
\end{align}
where we have defined
\begin{equation} \label{M_def}
    M_{kj} \equiv \frac{\Delta^6}{6912}\left[ 2 \nabla_k \nabla_l \bar u_m \nabla_j \nabla_m \bar u_l - \nabla^2 \bar u_l \nabla_k \nabla_l \bar u_j - \nabla_k \bar u_l \nabla^2 \nabla_l \bar u_j - 2 \nabla_k \nabla_m \bar u_l \nabla_l \nabla_m \bar u_j \right] + \mathcal{O} (\Delta^8)
\end{equation}
{\color{black}and 
\begin{equation} \label{M_def}
    F_{ij} = \frac{\Delta^6}{3456}\nabla_k\nabla^2 \bar u_{(i}  \nabla_k \nabla^2\bar f_{j)}+\mathcal{O}(\Delta^8)
\end{equation}
represents the effect of forcing. The explicit viscous correction is proportional to $\nu$ and is therefore small compared with the non-viscous residual terms in the high-Reynolds-number regime considered here. Since the evolution equation for $R_{ij}$ is not conservative, the asymptotic balance is not dominated by this viscous correction. Although this viscous contribution could in principle also be modeled, it is not expected to appear as a robust independent term in the present data-driven inference because the training data are in the high-Reynolds-number regime where this term is subdominant. We therefore do not retain it as a separate term in the closed model. By contrast, the tensor $M_{kj}$ survives in the high-Reynolds-number limit and cannot be analytically expressed in terms of $R_{ij}$ alone; hence its effect has to be modeled.}


{\color{black}Finally, note that the effect of forcing below the filter scale cannot be described using the formal moment expansion. Instead, the functional form of the forcing term in this case can be recovered using the definition \eqref{eq: R def} of $R_{ij}$. Specifically, we obtain
\begin{align}
    F_{ij}=\overline{f'_{(i}u'_{j)}}-\overline{f'_{(i}}\,\overline{u'_{j)}},
\end{align}
where, unlike $f'_i$, $u'_j$ can be expressed in terms of the resolved fields:
\begin{align}
    u'_j=-\frac{\Delta^2}{24}\nabla ^2\bar{u}_j+\mathcal{O}(\Delta^4).
\end{align}
Since, for small-scale forcing, $f'_i$ is not resolved, $F_{ij}$ has to be modeled as well.
}

\section{Sensitivity to the initialization of the Reynolds stress tensor}
\label{appendix:R_initialization}

\begin{figure}
    \centering
    \begin{subfigure}{0.45\textwidth}
        \centering
        \includegraphics[width=\textwidth]{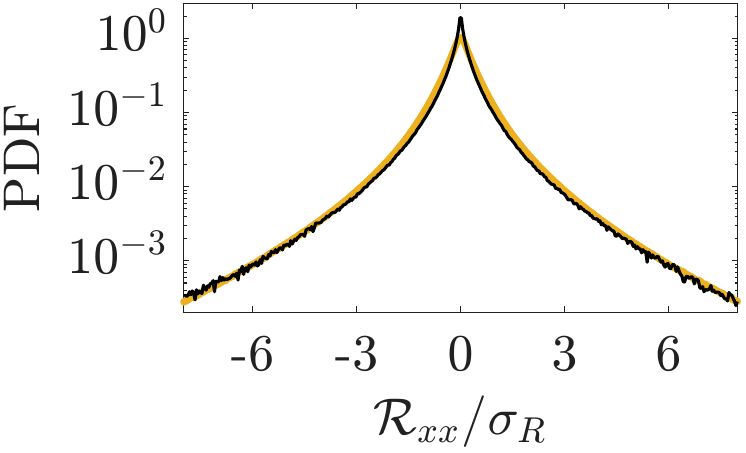}
        \caption{}
    \end{subfigure}
    \hspace{5mm}
    \begin{subfigure}{0.45\textwidth}
        \centering
        \includegraphics[width=\textwidth]{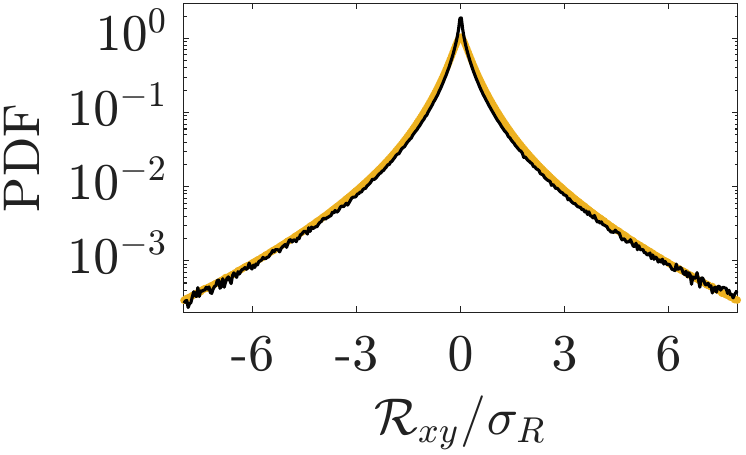}
        \caption{}
    \end{subfigure}

    \begin{subfigure}{0.45\textwidth}
        \centering
        \includegraphics[width=\textwidth]{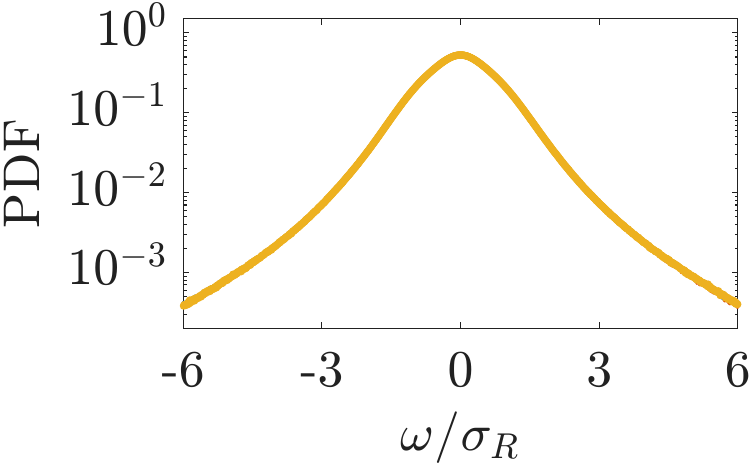}
        \caption{}
    \end{subfigure}
    \hspace{5mm}
    \begin{subfigure}{0.45\textwidth}
        \centering
    \includegraphics[width=\linewidth]{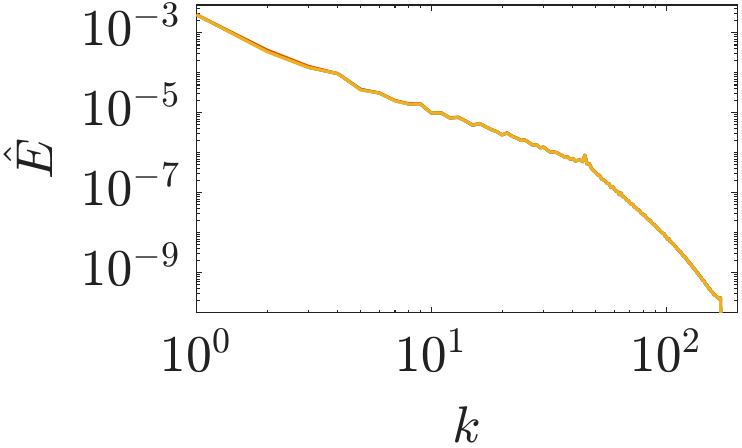}
        \caption{}
    \end{subfigure}
    \caption{Statistics for flow F5 obtained with NGMR using different initial conditions for $\mathcal{R}$. Panels (a) and (b) show the PDFs of the diagonal and off-diagonal components of $\mathcal{R}$, respectively. Panel (c) shows the corresponding vorticity PDF and panel (d) shows the power spectrum. The trivial initialization, shown in blue is compared with initializations based on Equations \eqref{leading_order_R} in orange and \eqref{eq: R def} in red. The black curves correspond to FDNS.
    }
    \label{fig:r_init_pdfs}
\end{figure}

{\color{black}
The NGMR model evolves an additional SGS tensor field $\mathcal{R}$ and therefore requires an initial condition for this field in online simulations. Since $\mathcal{R}$ represents SGS stresses, its exact initial value is generally not available in a practical LES calculation. It is therefore important to determine whether the long-time statistics predicted by NGMR depend sensitively on how $\mathcal{R}$ is initialized.
To test this, we compare three initializations for the forced flow F5. The first uses the formal definition of the Reynolds stress tensor, equation~\eqref{eq: R def}, and therefore requires DNS velocity data. The second uses the leading-order moment-expansion approximation, equation~\eqref{leading_order_R}, and therefore requires only filtered-DNS velocity data. The third uses the trivial initialization $\mathcal{R}_{ij}=0$, which requires neither fully resolved nor filtered velocity data.

For each initialization of $\mathcal{R}$, the NGMR model is evolved, using the same setup and parameters as described in Section \ref{section:forced}, until the initial transient has decayed and the flow has reached the statistically stationary regime. We then compare the probability density functions of the two independent components of the traceless Reynolds stress tensor, $\mathcal{R}_{xy}=\mathcal{R}_{yx}$ and $\mathcal{R}_{xx}=-\mathcal{R}_{yy}$. As shown in \autoref{fig:r_init_pdfs}(a-b), these PDFs are indistinguishable and also agree closely with the corresponding FDNS statistics. The statistics of other resolved quantities for different initializations are also indistinguishable. 
In particular, the PDFs of vorticity and the time-averaged power spectra are compared in \autoref{fig:r_init_pdfs}(c-d). The spectra for the three initializations lie essentially on top of one another across the resolved wavenumber range, and also closely follow the FDNS spectrum. These results show that, for the statistically stationary forced flow considered here, the long-time NGMR statistics are insensitive to the initialization of $\mathcal{R}_{ij}$. 
}

\bibliographystyle{jfm}
\bibliography{jfm}

\end{document}